\documentclass[journal,12pt]{IEEEtran}
\usepackage{graphicx}

\usepackage{epsfig}
\usepackage{times}
\usepackage[nocompress]{cite}
\usepackage{amsmath,amssymb}
\usepackage{pifont}
\usepackage{balance,simplemargins}
\usepackage{graphicx}
\usepackage{epsf}
\usepackage{xcolor}
\usepackage{simplemargins}
\setallmargins{1in}
\settopmargin{19mm}
\setbottommargin{43mm}
\setrightmargin{12mm}
\setleftmargin{12mm}

\begin{document}
\title{Design of a Scalable DNA Shearing System Using Phased-Array Fresnel Lens Transducers
\author{Kapil~Dev$^\dagger$, Smriti~Sharma$^\ddagger$, Vibhu~Vivek$^\ddagger$, Babur~Hadimioglu$^\ddagger$, and Yehia~Massoud$^\dagger$\\
$^\dagger$Department of Electrical and Computer Engineering, University of Alabama at Birmingham, Birmingham, Al 35205\\
$^\ddagger$Microsonic Systems San Jose, CA 95134; vibhu.vivek@microsonics.com}}
\maketitle


\begin{abstract}
In this paper, we present the design of a Deoxyribonucleic Acid (DNA) shearing system based on unique acoustic waves generated using a phased-array Fresnel Lens transducer. Four $90^{\circ}$ sector-transducers are used to build a circular array transducer. Acoustic simulation results for particle displacement are provided for cases when one, two, three, and all four transducers in the array are excited with RF signals. Each $90^{\circ}$-transducer is excited with separate RF signal of same or different phase. The proposed transducer structure generates bulk lateral ultrasonic waves in the DNA sample; the lateral waves produce both convergence and vortexing effects in the sample. The converged lateral acoustic waves are required to break the DNA sample-meniscus inside the tube and the rotational component of acoustic field is used to recirculate the DNA sample to get homogeneous shearing with tight fragment distribution. Finally, we present the experimental results of DNA sheared to different mean fragment sizes using the proposed system and validate the shearing-capability of the proposed system.
\end{abstract}


\section{Introduction}
\label{sec:intro}
With the continuous research and advances in Deoxyribonucleic acid (DNA) sequencing technologies the need for an efficient DNA shearing system has increased more than ever before. There are different shearing techniques used by researchers and genomic institutes; among them include enzymatic digestion~\cite{NEBioLabs}, nebulization~\cite{Invitrogen}, hydroshear~\cite{JGI,DigiLab} and sonication based methods~\cite{Larguinhoa2010, ElsnerDNA1989, MSSpatent2012_WO, MSSpatent2012_EP, MSSpatent2012_JP, MSSpatent2012_US, kdevICES2011-2, kdevMSthesis2011, dev2019analytical_Arxiv, kdevICM2019}.  The enzymatic digestion is good for creating fixed-size DNA fragments, e.g. DNA markers. Since the sequencing algorithms require the DNA fragments of somewhat random fragment sizes, the enzymatic digestion technique is not good for sequencing applications. Further, the enzyme based DNA shearing method is somewhat slow and is not capable to meet the high-throughput needs of next generation DNA sequencing systems. Nebulization can provide random fragments of DNA for sequencing but has specific requirements for the volume and concentration of input DNA~\cite{Lucigen}. Also, there are disadvantages in terms of broad distribution of fragments, loss of sample due to atomization, and a risk of DNA cross-contamination in the nebulization process. Hydroshear can provide size-specific random DNA fragments with minimal DNA damage suitable for sequencing strategies. But the hydroshear equipment is quite complicated and the complete process is very time consuming. Sonication is the only powerful and controlled method which can meet the throughput demands of next generation DNA sequencing technologies being developed by companies like Illumina, Pacific Biosciences, InteGenX, JGI Sanger, Roche 454, Applied Biosystems Inc., Helicos Biosciences Corp., RainDance Technologies, Microsonic Systems, etc. Ultrasonic probe, sonoreactor, ultrasonic bath, and adaptive focused acoustic (AFA) are some of the current ultrasound-based platforms for DNA shearing~\cite{Larguinhoa2010, MannBioSens2004}. Ultrasonic probe has the risk of contamination in carry over DNA sample, sonoreactor generates broad size-distribution and hence lot of DNA goes waste during the size-selection process before sequencing. AFA based shearing from Covaris Inc. is an efficient shearing technique but it may be difficult to scale their system for high-throughput shearing~\cite{Covaris}. 

\begin{figure*}[tb]
\begin{center}
\begin{tabular}{cc}
\includegraphics[width=0.3\textwidth]{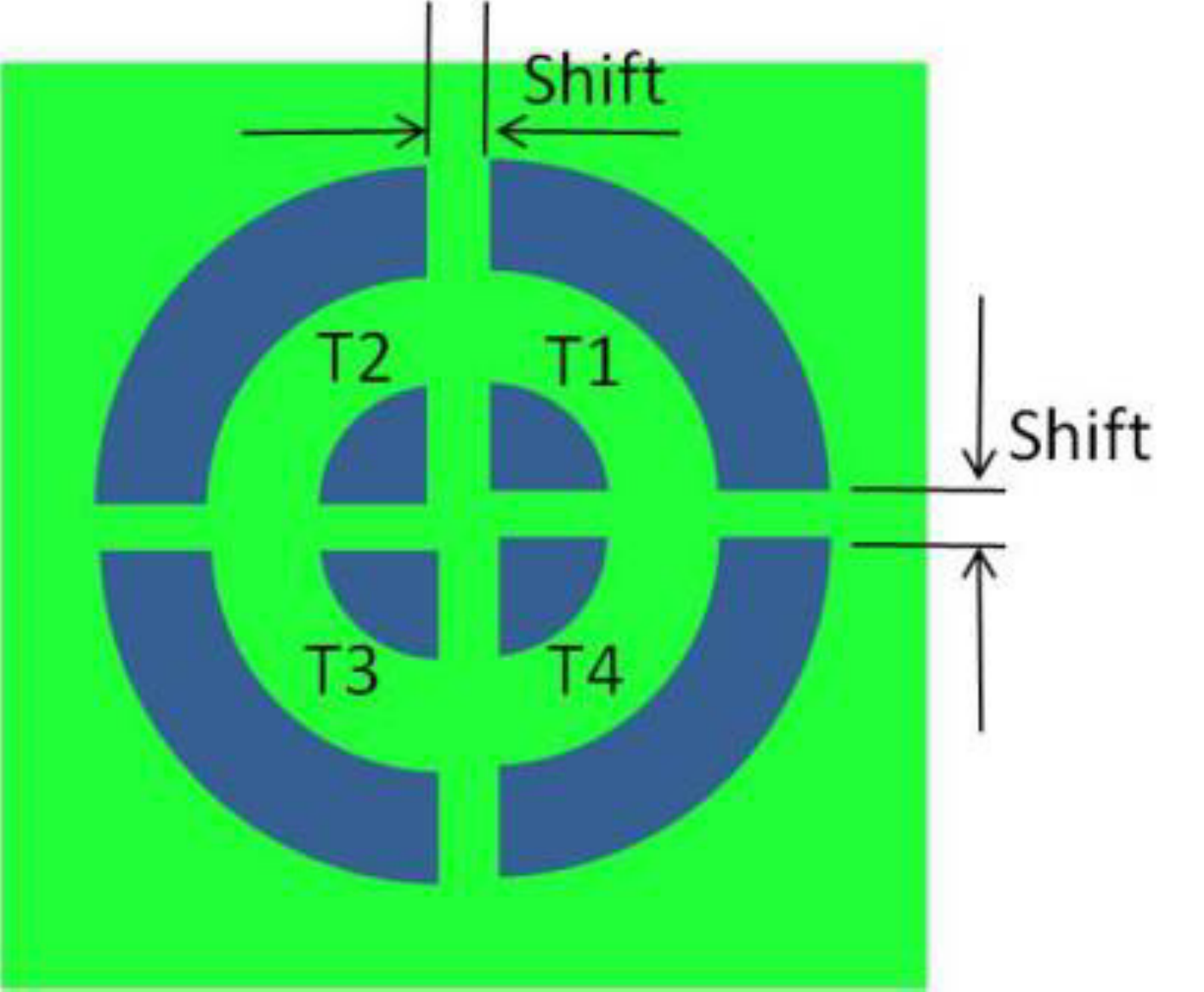} &
\hspace{0.6in}
\includegraphics[width=0.5\textwidth]{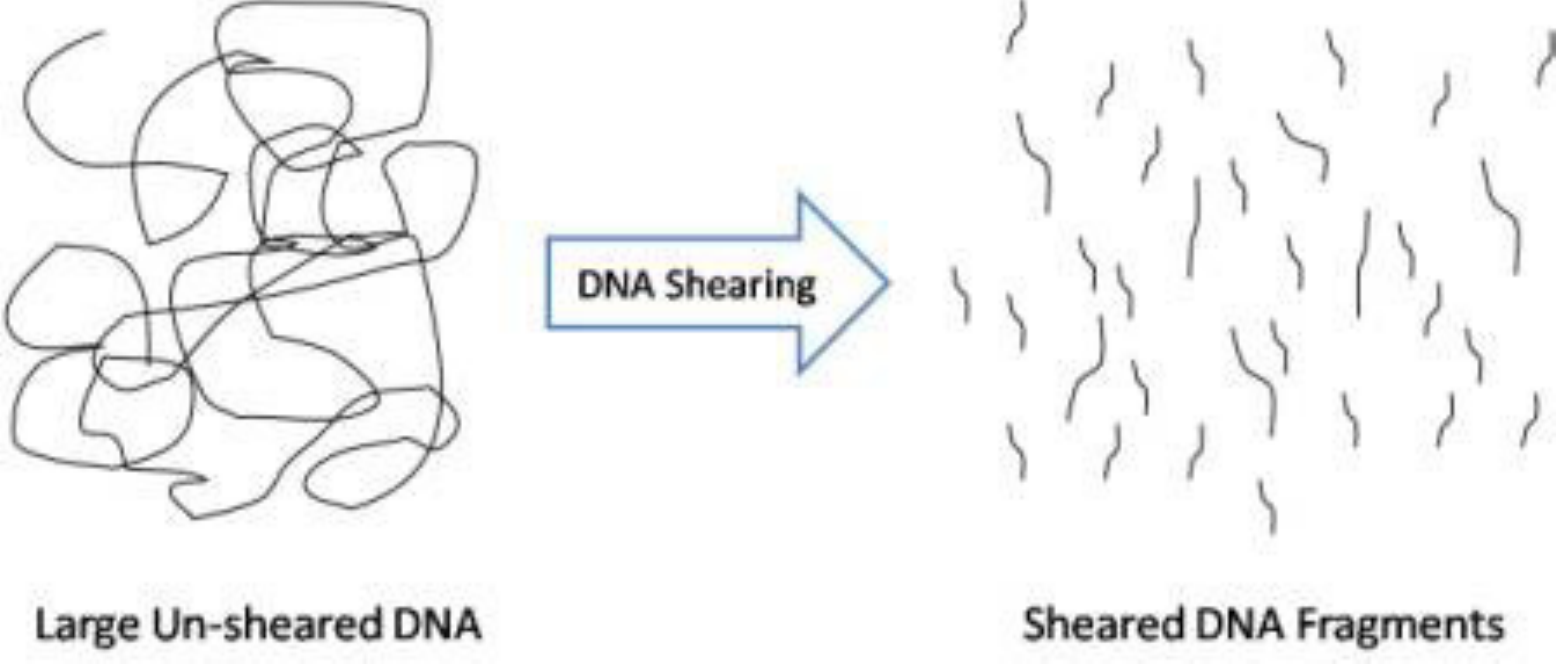}\\
(a) & (b)\\
\end{tabular}
\end{center}
\caption{ (a) Proposed array of $90^{\circ}$ transducer structure; (b) Un-sheared and sheared DNA.} 
\label{fig:TransducerArray_n_LargeSmallDNA}
\vspace{-0.2in}
\end{figure*}

\begin{figure}[b]
\centering 
\includegraphics[width=0.94\columnwidth]{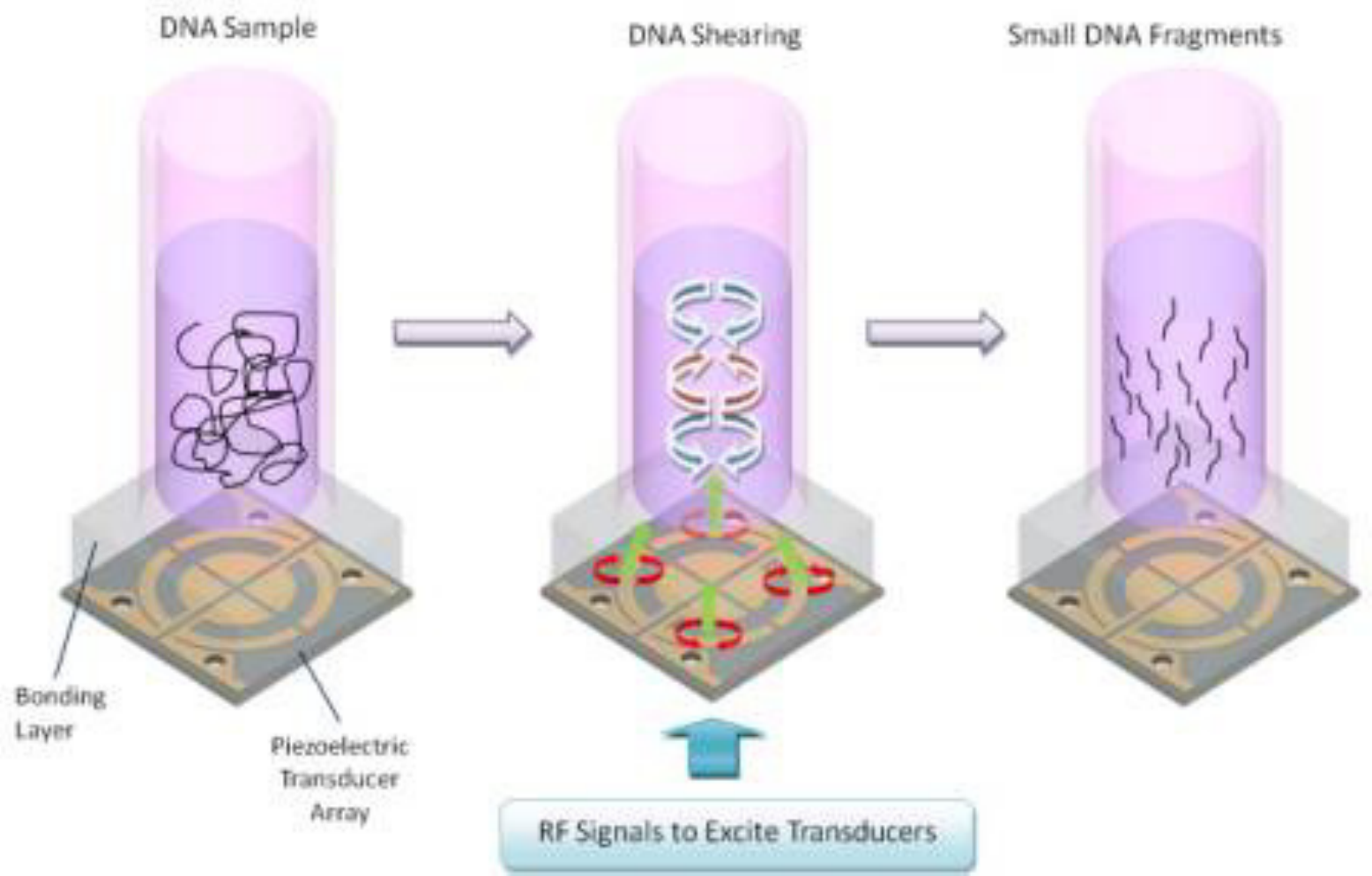}
\caption{DNA shearing using lateral shear waves}
\label{fig:DNAshearingBlockDiag}
\vspace{-0.2in}
\end{figure}

In this paper, we propose the design of a fast, efficient, controlled, and scalable DNA shearing system that uses uniquely generated ultrasonic shear forces to shear the DNA to desired fragment sizes. The proposed system uses a specially designed piezoelectric transducer to generate the shear forces. The piezoelectric transducer has $90^{\circ}$ sectored annular electrodes that generates lateral acoustic waves in the space over it when an RF excitation is applied across the transducer. We arrange four $90^{\circ}$ sectored transducers in a circular array as shown in figure~\ref{fig:TransducerArray_n_LargeSmallDNA} and excite each transducer with dedicated RF signal having different phase and duty cycle. The acoustic field generated from the phased-array transducer structure has both converging and vortexing effects, shown in figure~\ref{fig:DNAshearingBlockDiag}. We have successfully used such a transducer structure to shear the DNA to different mean fragment sizes~\cite{kdevICES2011-1}; this manuscript describes the theory and experiments used to build the scalable shearing system in greater detail. 

The rest of the paper is organized as follows. Section~\ref{sec:PhaseArraySectoredXducer} describes the proposed array-transducer which is used to shear the DNA. In this section, we also discuss the acoustic wave patterns generated in space when the transducers are excited by RF signals of different phase and duty-cycles. Simulation results are presented for all 5 cases- when one, any two, any three, and all four transducers are active in a circular array of 4 transducers.   Section~\ref{sec:XperimentResults} presents the results of shearing experiments we performed using our system. Finally, we conclude the paper in section~\ref{sec:conclusion}.

\section{Phased-Array Piezoelectric Transducer}
\label{sec:PhaseArraySectoredXducer}
It was first presented in~\cite{VibhuMEMS2000} that a $90^{\circ}$ sector-annular piezoelectric transducer could be used to generate bulk lateral acoustic waves which can mix, thaw and solubilize the microfluidic samples without degrading or denaturing the samples. We use $90^{\circ}$ sector-transducer as the basic unit and arrange four $90^{\circ}$ transducers, T1-T4, in a circular fashion to achieve both converging and vortexing effects. Such a transducer design is shown in figure-\ref{fig:TransducerArray_n_LargeSmallDNA} (a). Depending on the value of shift between two transducers the transducer has localized or distributed maximum acoustic field in space.


\begin{figure*}[t]
\begin{center}
\begin{tabular}{ccc}
\includegraphics[width=0.28\textwidth]{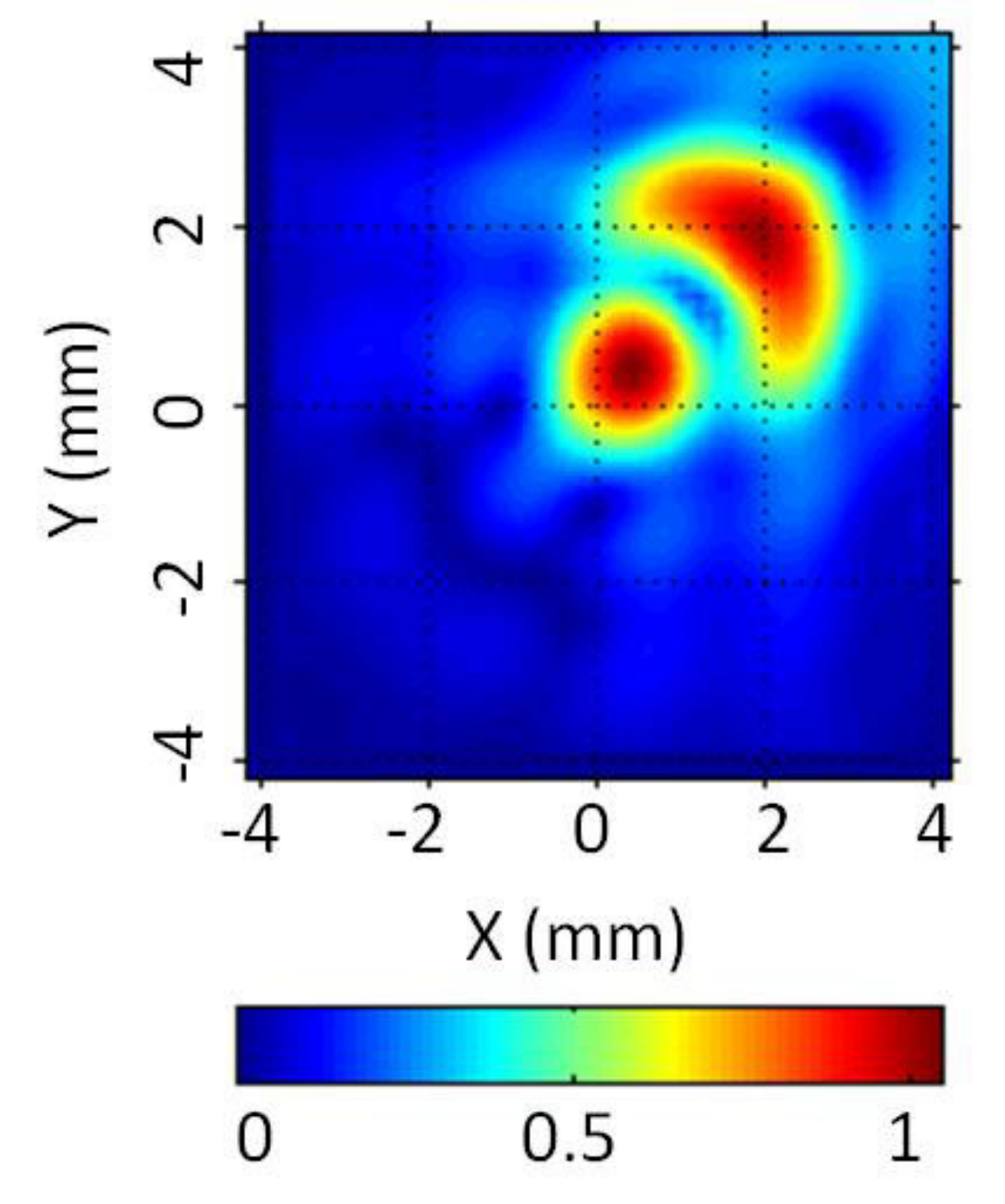} &
\includegraphics[width=0.28\textwidth]{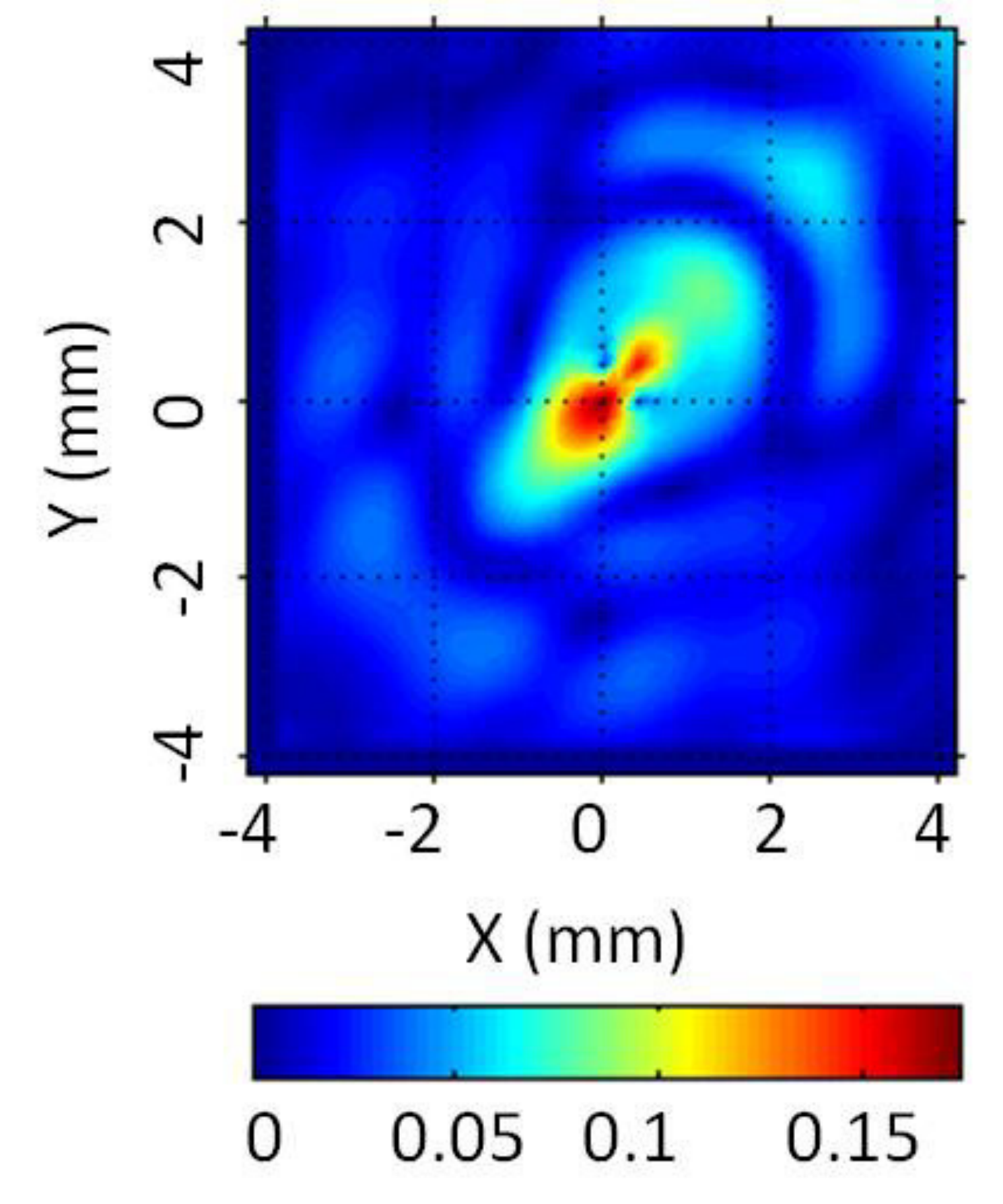} &
\includegraphics[width=0.28\textwidth]{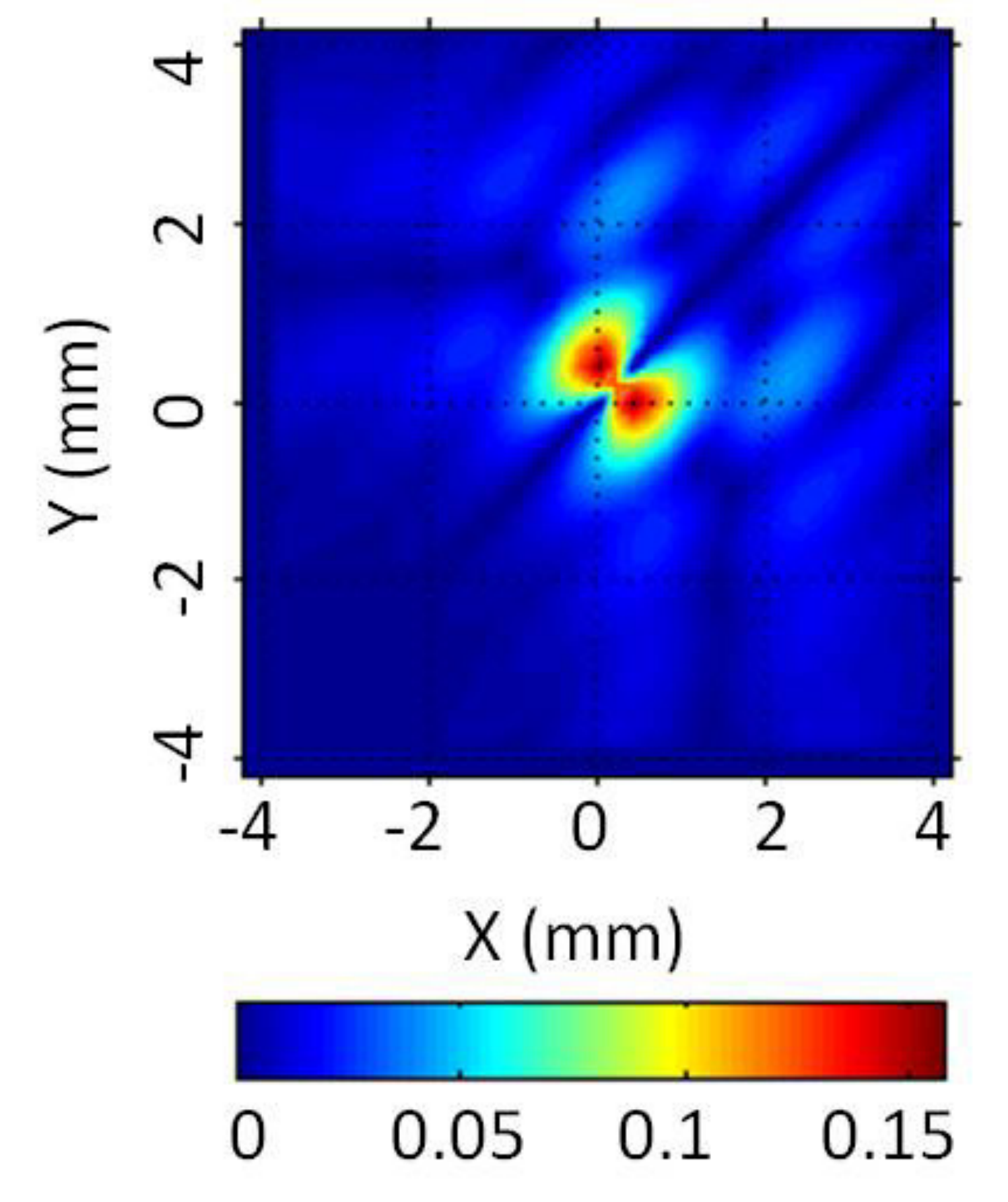}\\
(a) & (b) & (c)\\
\end{tabular}
\end{center}
\vspace{-0.1in}
\caption{Particle displacement components when one  transducer (T1) is ON: (a) $u_z$, (b) $u_r$, and (c) $u_\psi$.} 
\label{fig:T1OnOffCases}
\vspace{-0.22in}
\end{figure*}

\noindent 
{\bf{Computing particle displacement from phased-array transducers.} }
Assuming the transducer array is placed at the origin, the displacement potential at any point P($r, \psi,z$) due to the transducer array is computed using full Rayleigh -Sommerfeld integral~\cite{GSKinoBook1987, RosenbaumBAWbook1988}. 

\begin{equation}
\label{eq:PhiBasic}
\Phi\left(r,\psi,z\right) = \sum_{i=1}^{4} -\frac{u_i(t)}{2\pi} \int_{S_i} \frac{e^{-\left(\alpha +jk \right)R}} {R} dS_i
\end{equation}

Where, i=\{1, 2, 3, 4\} denotes transducers T1, T2, T3 and T4 respectively, $u_i(t)$ is the particle displacement right above the surface of $i^{th}$ $90^{\circ}$ sector-transducer and it is the function of time having same time varying properties as the applied RF signal, $\alpha$ is the acoustic attenuation constant of the medium, $k$ is the wave vector ($= 2\pi/\lambda$) and $R$ is the distance between point P and the elemental area $dS_i$. The wavelength ($\lambda$) of acoustic wave in the medium is calculated from the frequency of RF signal and the speed of sound in the medium. At a time any one or more transducers could be active and there may be a phase shift between the RF signals applied to these transducers. In equation~\ref{eq:PhiBasic}, $u_i(t)$ incorporates the properties of RF signal applied to $i^{th}$ transducer. We use superposition to calculate the resultant acoustic potential at a point due to different active transducers by taking in to account the shift between them. 

Once the acoustic potential is computed at any point, the relative particle displacement (in radial ($u_r$), vertical ($u_z$), and circumferential ($u_\psi$) directions) are calculated by differentiating the acoustic potential at that point. That is:

\begin{equation}
\label{eq:u_vs_Phi}
u = \nabla \Phi \left( r, \psi, z \right) = \sum_{i=1}^{4} \left( \frac{\partial} {\partial r} \hat{r} + \frac{\partial} {r\partial \psi} \hat{\psi} + \frac{\partial} {\partial z} \hat{z} \right)
\end{equation}

In our transducer design, each sector-transducer is excited by a separate RF amplifiers. The RF signals applied to $i^{th}$ transducer is $RF_i$ and is given by equation~(\ref{eq:RFsignals}):
\vspace{-0.1in}
\begin{equation}
\label{eq:RFsignals}
RF_i = D_i(t)V_isin\left(\omega t+p_i \right)
\end{equation}

Where, i=\{1, 2, 3, 4\} denotes the index of transducers \{T1, T2, T3, T4\}, $D_i(t)$, $V_i$ and $p_i$ are the duty cycle, peak voltage and phase of the RF signal applied to $i^{th}$ transducer. Having independent RF excitation for each transducer gives us the flexibility to switch-on and -off the sector-transducers independently which helps in efficiently recirculating the DNA sample during shearing process. Typically, we keep the frequency, repetition rate (included in $D_i(t)$ term) and the voltage of RF signal same for all the sector-transducers in the array. We have used the separation between any two adjacent transducers approximately equal to $\lambda /2$ (``shift"=0.2 mm); the value is chosen so that we can fit four sector-transducers in the 9mm by 9mm footprint, which is the pitch of a standard 96-well microplate.

\begin{figure*}[t]
\begin{center}
\begin{tabular}{ccc}
\includegraphics[width=0.28\textwidth]{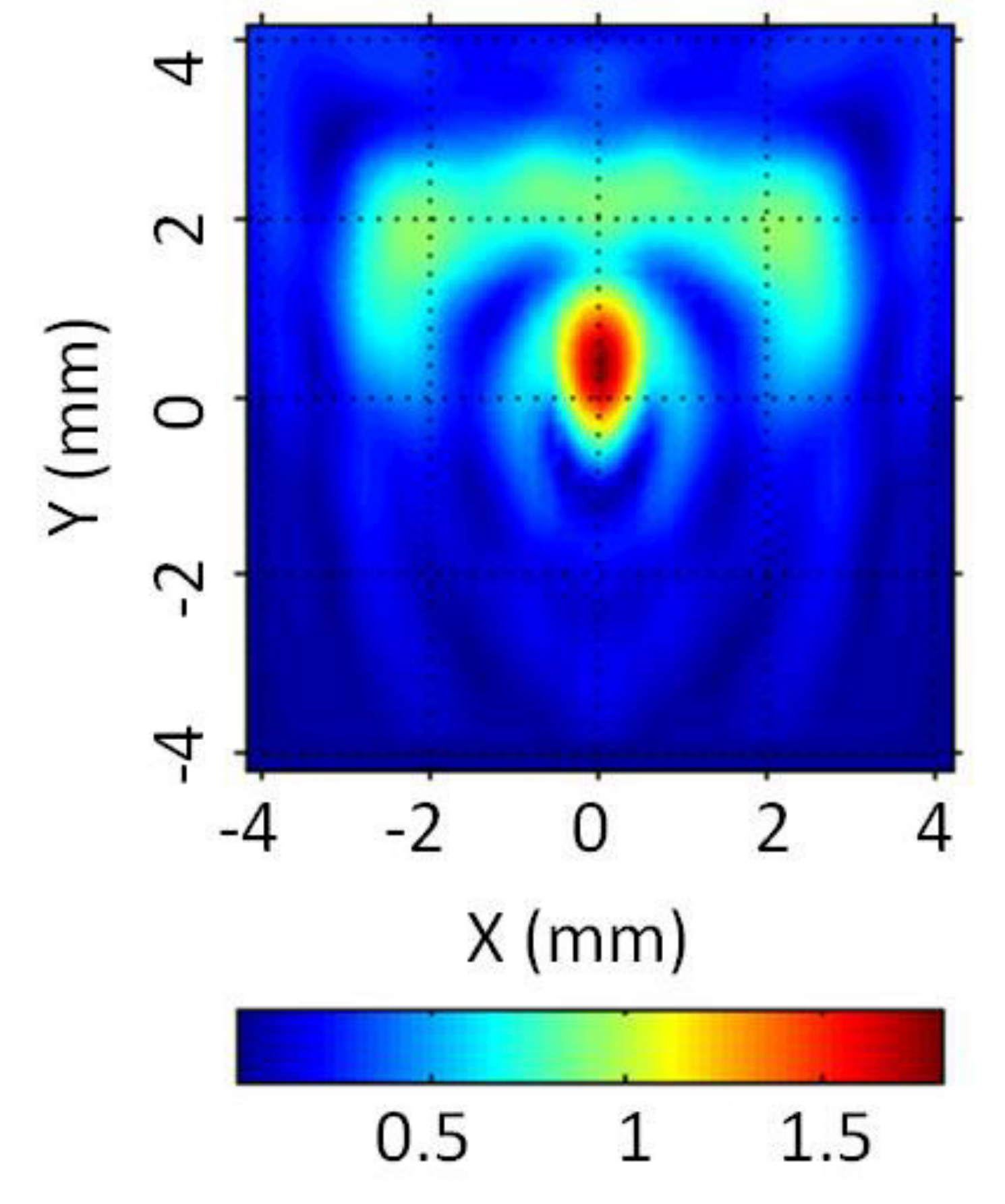} &
\includegraphics[width=0.28\textwidth]{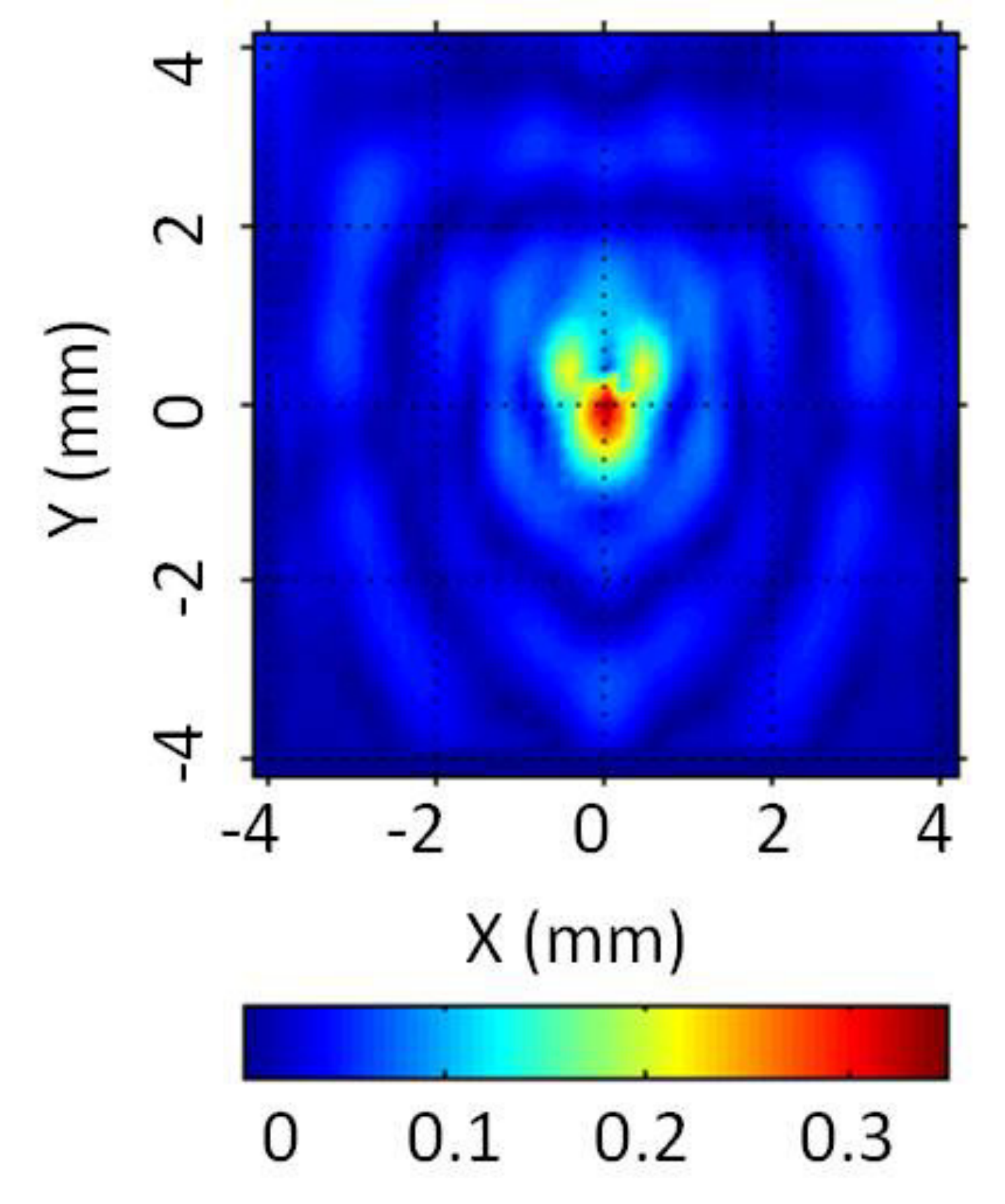} &
\includegraphics[width=0.28\textwidth]{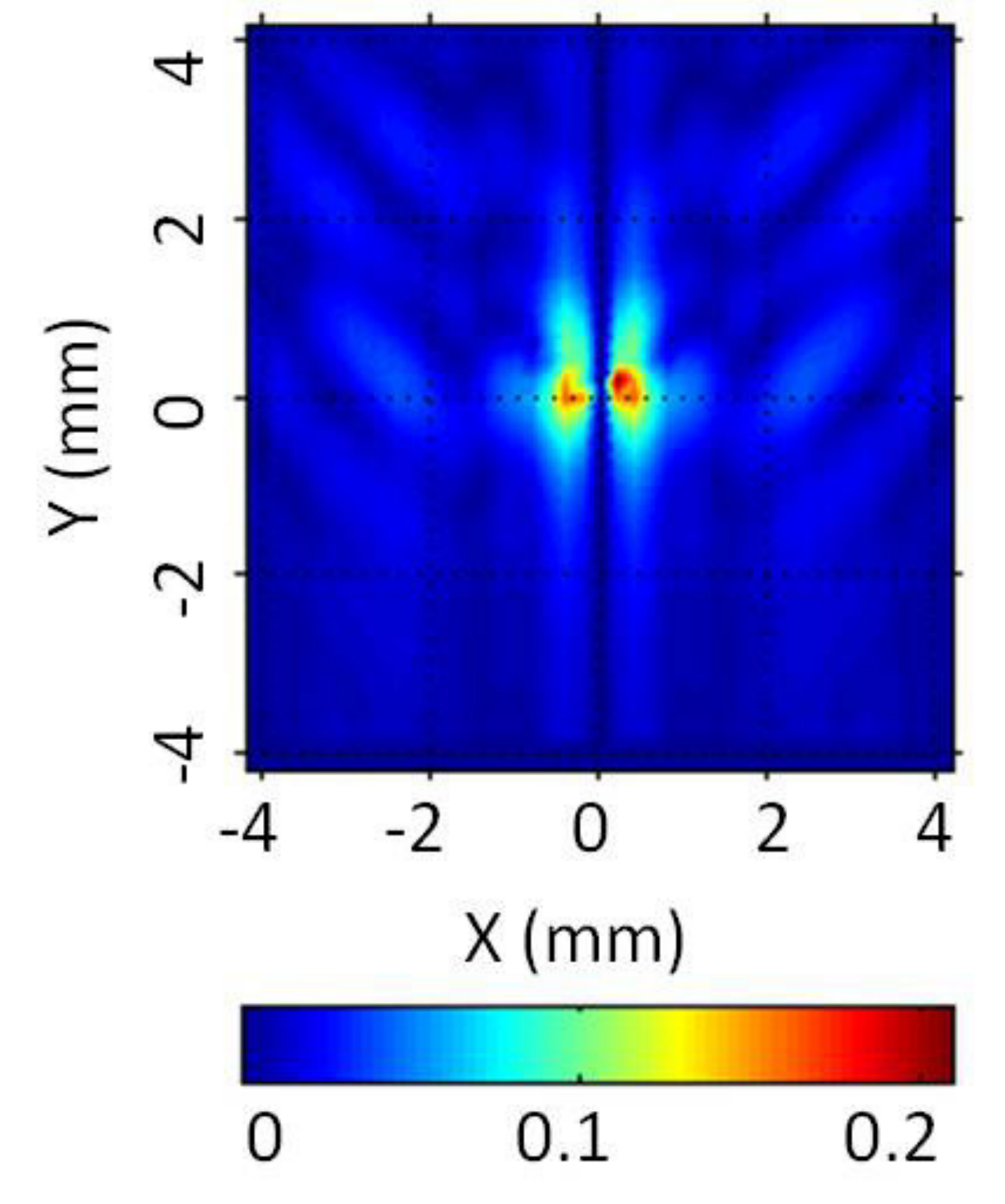}\\
(a) & (b) & (c)\\
\includegraphics[width=0.28\textwidth]{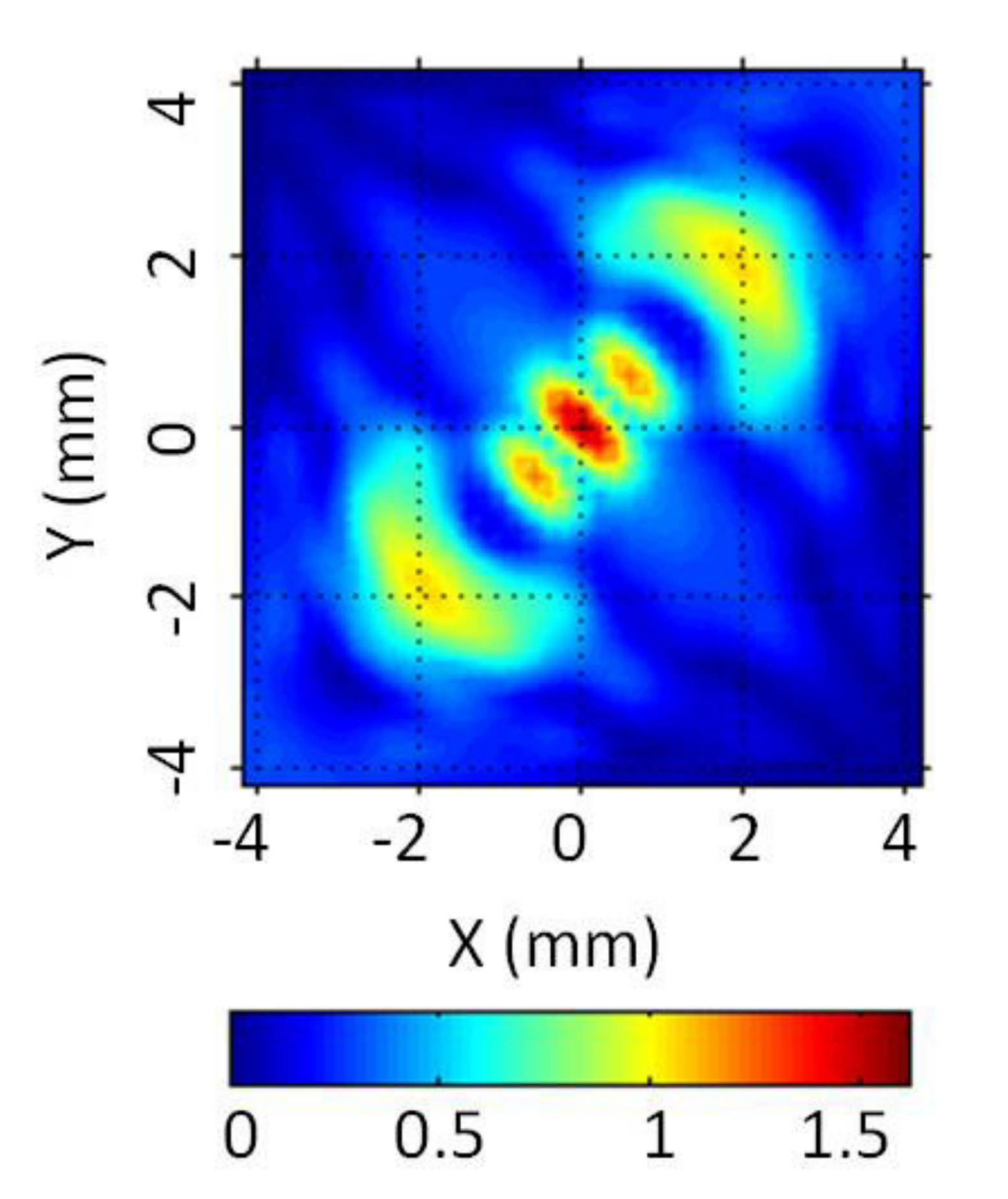} &
\includegraphics[width=0.28\textwidth]{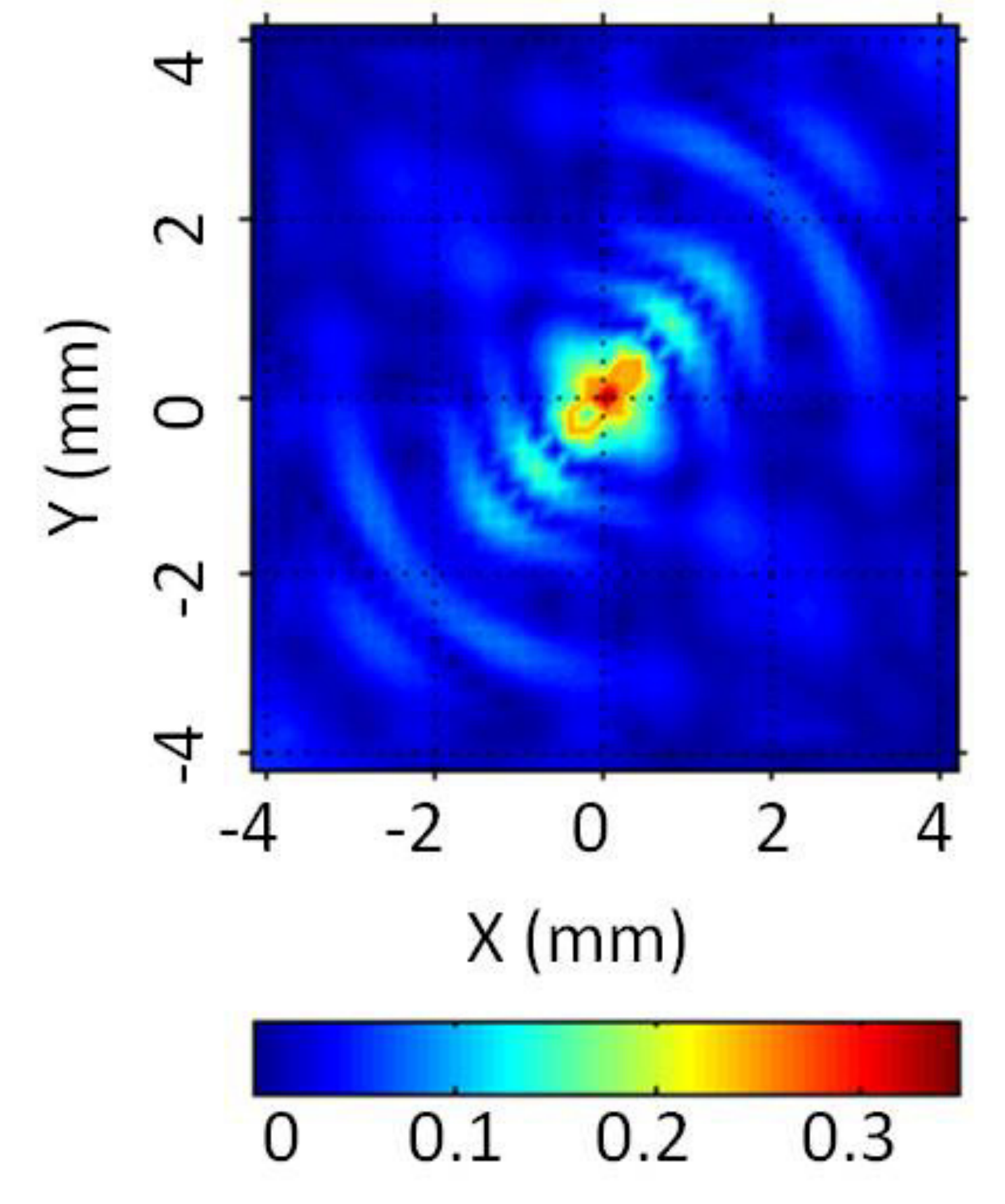} &
\includegraphics[width=0.28\textwidth]{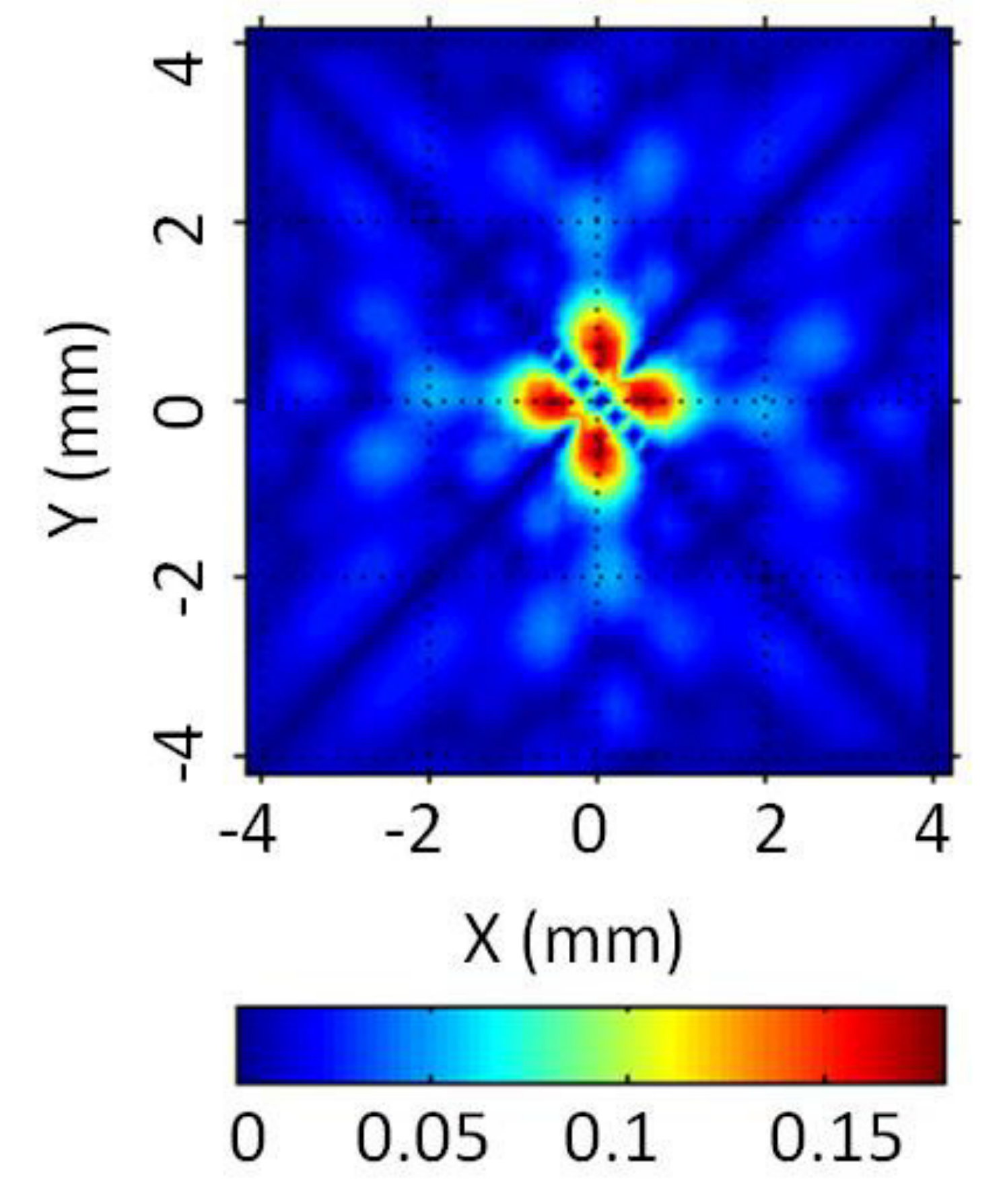}\\
(d) & (e) & (f)\\
\end{tabular}
\end{center}
\vspace{-0.1in}
\caption{Particle displacement components when two transducers are ON: (a) $u_z$, (b) $u_r$, and (c) $u_\psi$, when T1 and T2 (adjacent transducers) are active; (d) $u_z$, (e) $u_r$, and (f) $u_\psi$, when T1 and T3 (diagonally opposite transducers) are active} 
\label{fig:T12T13OnOffCases}
\vspace{-0.22in}
\end{figure*}

\begin{figure*}[t]
\begin{center}
\begin{tabular}{ccc}
\includegraphics[width=0.28\textwidth]{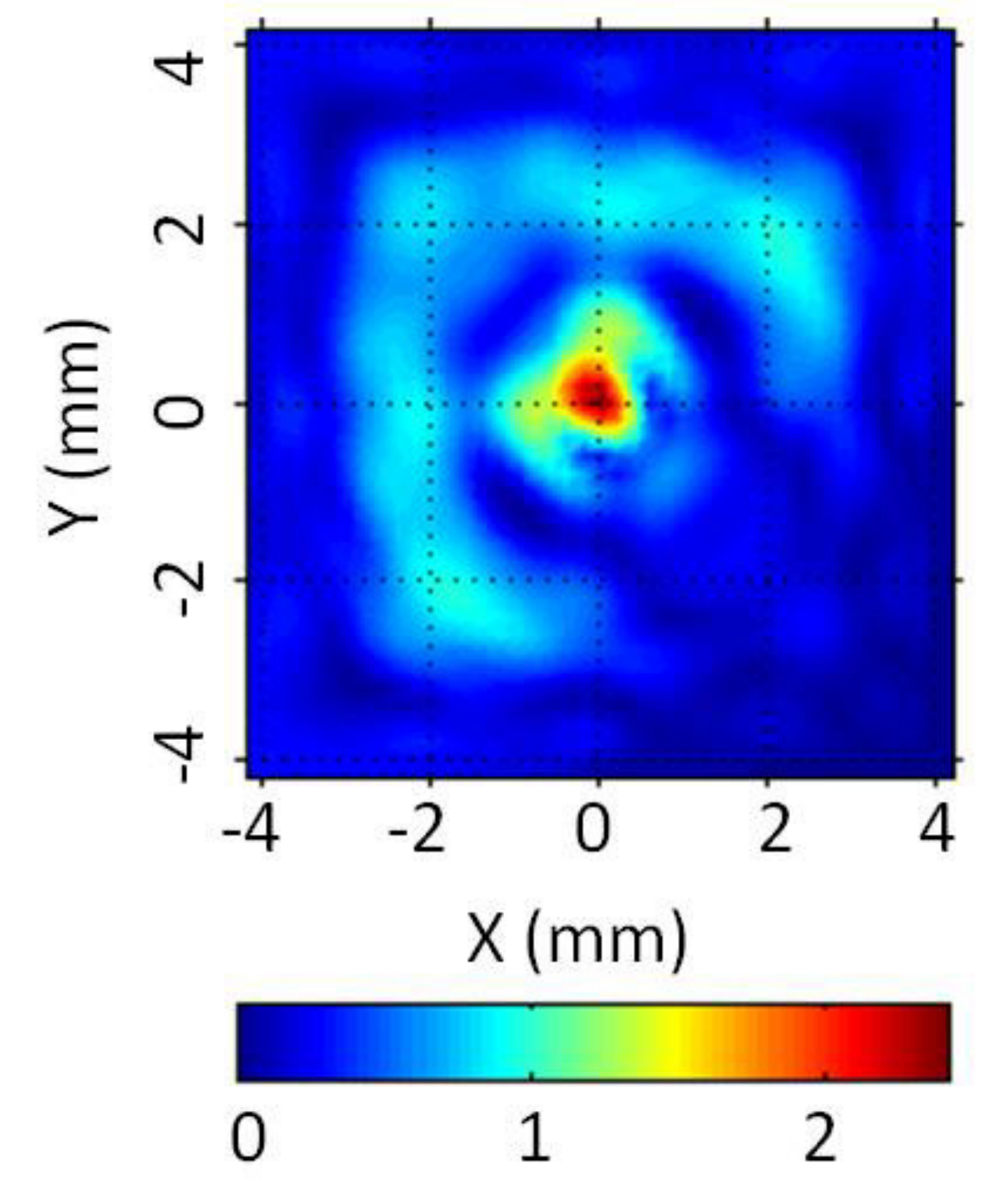} &
\includegraphics[width=0.28\textwidth]{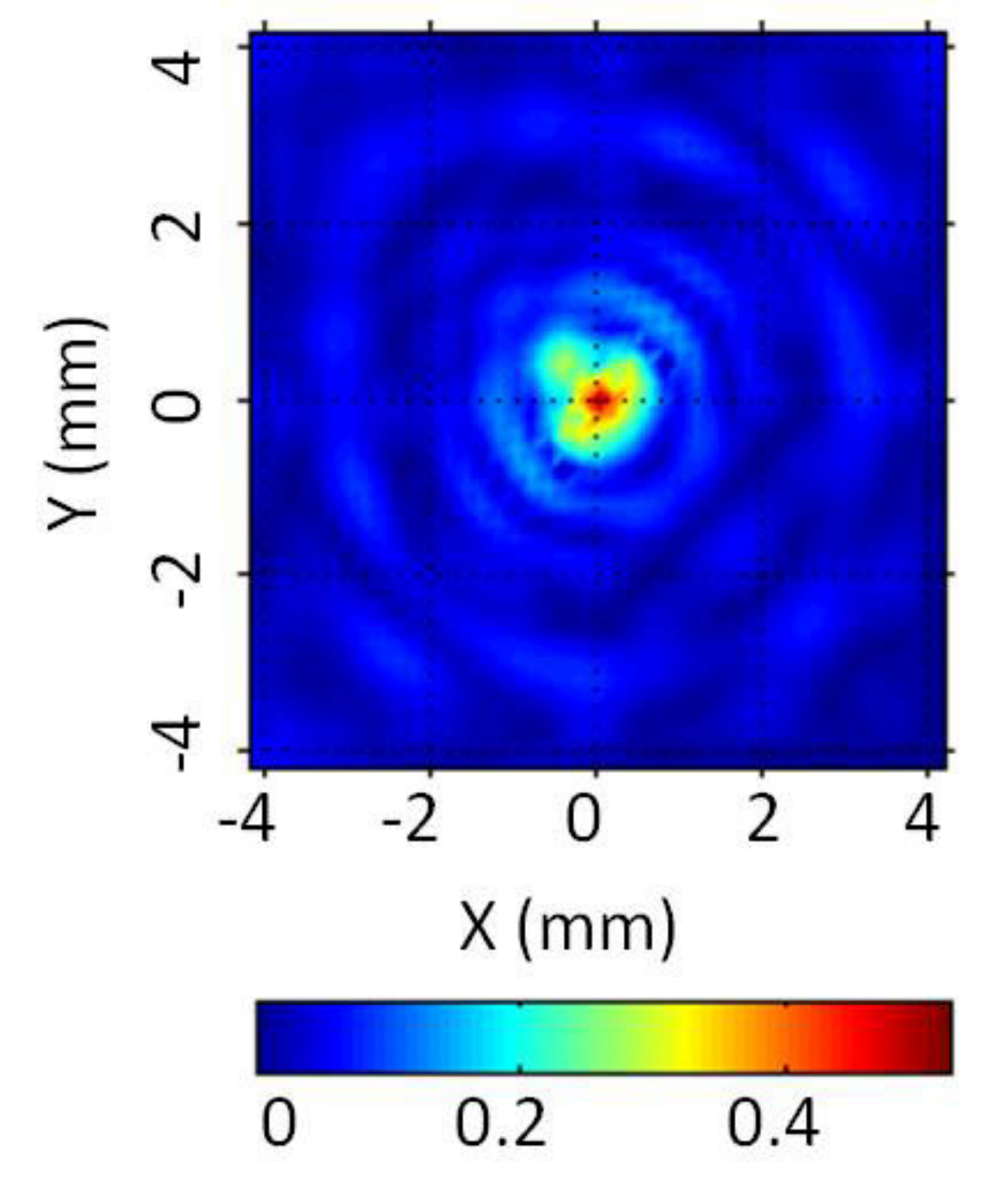} &
\includegraphics[width=0.28\textwidth]{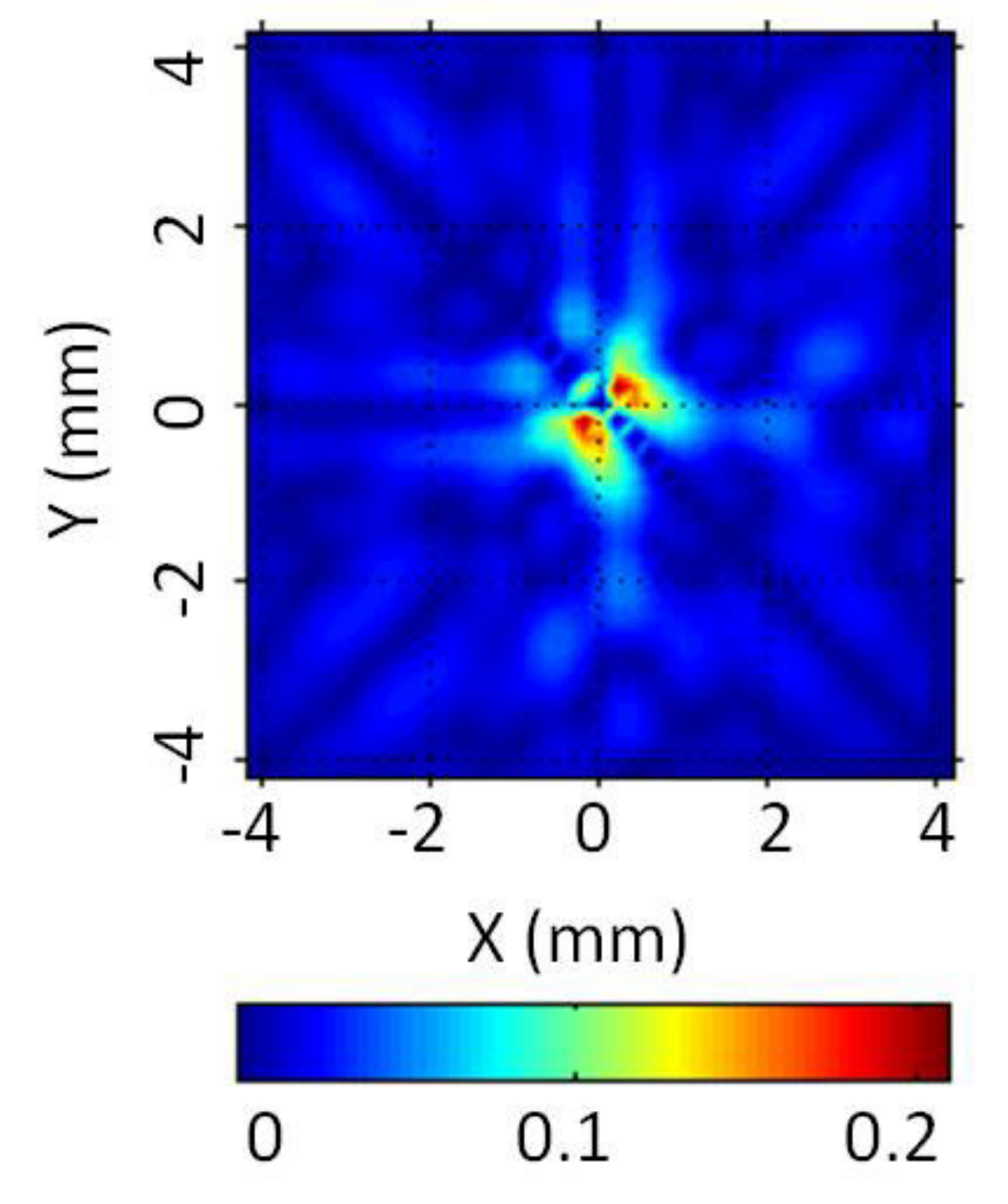}\\
(a) & (b) & (c)\\
\end{tabular}
\end{center}
\vspace{-0.1in}
\caption{Particle displacement components when three transducers (T1, T2, T3) are active: (a) $u_z$, (b) $u_r$, and (c) $u_\psi$} 
\label{fig:T123OnOffCases}
\end{figure*}

\noindent
{\bf{Exciting phased-array transducers.} }
With four sector-transducers (T1-T4) in the array, there could be total five types of acoustic wave patterns generated in water medium depending on the number of active transducers at a particular instant of time. These five cases are as discussed below. 
\begin{enumerate}
\item \textit{Only one transducer is active}: In this case only one transducer is active out of T1-T4 transducers at a particular instant of time. We computed the particle displacement at 13mm height above the transducer surface when only one transducer is active. The vertical, radial and circumferential component of particle displacements when only T1 transducer is active is shown in figure-\ref{fig:T1OnOffCases}(a)-(c). It could be noticed from figure-\ref{fig:T1OnOffCases}(a) that the vertical component of particle displacement has its maxima located in the same quadrant in which the transducer is located. As expected, the radial component has its maxima located in the diagonally opposite quadrant; for example, when transducer T1 (placed in first quadrant) is active, the radial component is directed towards the third quadrant as shown in figure-\ref{fig:T1OnOffCases}(b). Further, from figure-\ref{fig:T1OnOffCases}(c), we observe that the rotational component of acoustic particle displacement has its peaks intensity points located in the plane perpendicular to the orientation of the transducer. The maximum intensity lobes of $u_\psi$ for transducer T1 (placed in first quadrant with $45^{\circ}$ orientation) are in the plane aligned in ($135^{\circ}$, $315^{\circ}$) directions. The orientation of the plane in which the maximum intensity of $u_\psi$ lies depends on the start and end angle of the sectored transducer. Also, it is important to mention that the two lobes of $u_\psi$ have rotational effects in opposite directions, which is utilized to homogenize the DNA sample for uniform DNA shearing. 

\item \textit{Any two adjacent transducers are active}: In this case, any two \emph{adjacent} transducers are active at a particular instant of time. The possible cases are: (T1,T2) or (T2,T3) or (T3,T4) or (T1,T4) active at a time; they are all symmetric cases. The simulation results for particle displacements ($u_z$, $u_r$, $u_\psi$) when transducers T1 and T2 are active are shown in figure-\ref{fig:T12T13OnOffCases}(a)-(c). Following the same explanation as provided in the previous case, the maximum intensity points of $u_z$ are mainly located in the first and second quadrant because the active transducers are located in the first and second quadrants (figure-\ref{fig:T12T13OnOffCases}(a)). Since the transducer-array consisting of T1 and T2 transducers has $90^{\circ}$ placement orientation, the radial component ($u_\psi$) has its maxima located on the -ve y-axis, diagonally opposite to the orientation of two active transducers, as shown in figure-\ref{fig:T12T13OnOffCases}(b). As we could see in figure-\ref{fig:T12T13OnOffCases}(c), the maximum intensity of $u_\psi$ in this case lies in the plane aligned in ($0^{\circ}$, $180^{\circ}$) directions. It is also important to mention that the absolute intensity of $u_z$, $u_r$, and $u_\psi$ are higher in this case as compared to the previous case, where only one transducer is active, because the area of active source of ultrasonic waves is higher in this case.

\item \textit{Any two diagonal transducers are active}: Similar to the previous case, in this case also only two transducers are active at a time. However, unlike previous case, the two active transducers are diagonally opposite ones in the array of four transducers, that is (T1,T3) or (T2,T4). As we can observe from the simulation results shown in figure-\ref{fig:T12T13OnOffCases}(d)-(f), the maximum intensity of $u_z$ lies in the quadrants where the active transducers are situated (quadrant I and III when transducers T1 and T3 are ON). The radial component ($u_r$) is maximum at center because the two diagonally opposite transducers would add up to maximum at center. This is unlike in case-1, where the $u_r$ has its maximum intensity in the III quadrant when T1 is active. Further, from figure-\ref{fig:T12T13OnOffCases}(f), we notice that the rotational component of particle displacement is having four peak lobes in four directions corresponding to the start and the end angle of two active transducers, that is, towards $0^{\circ}$ and $90^{\circ}$ due to T1 and towards $180^{\circ}$ and $270^{\circ}$ due to T3. So, we have four vortexing loci in the DNA sample tube in this case which could be used to uniformly mix the DNA sample during shearing process.
 
\item \textit{Any three transducers are active}: In this case, any three out of four transducers are active at a particular instant of time. For example when T1, T2, and T3 are active, the maximum intensity of $u_z$ lies in the first three quadrants, maxima of $u_r$ is in the fourth quadrant (opposite to the orientation of three active transducers and maximum $u_\psi$ points are in the first and third quadrants, close to the start-angle ($0^{\circ}$) and the end-angle ($270^{\circ}$) of three-transducer array. The simulation plots for different components of particle displacement ($u_z$, $u_r$, and $u_\psi$) are shown in figure-\ref{fig:T123OnOffCases}(a), (b), and (c) respectively.

\begin{figure*}[t]
\begin{center}
\begin{tabular}{ccc}
\includegraphics[width=0.28\textwidth]{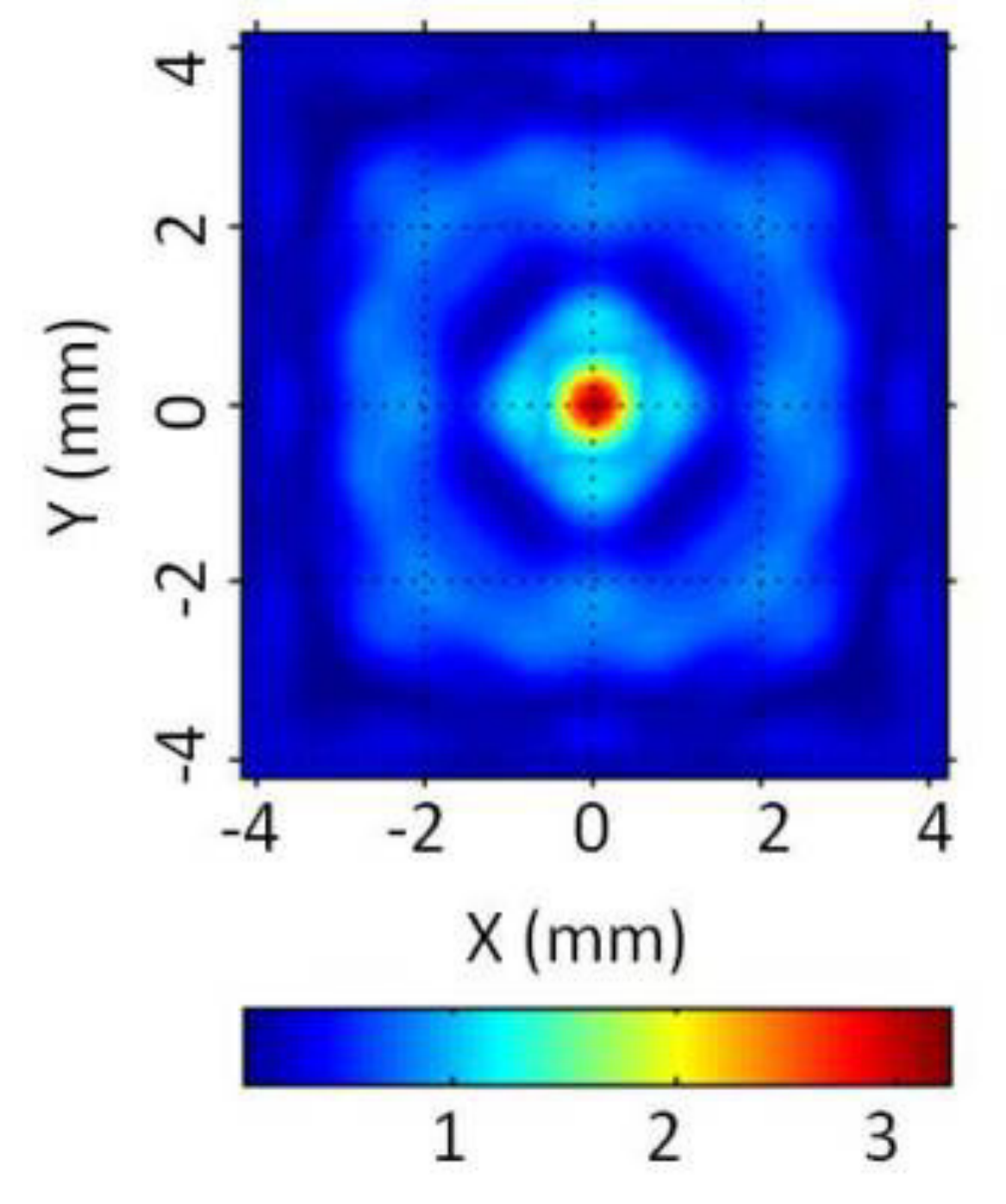} &
\includegraphics[width=0.28\textwidth]{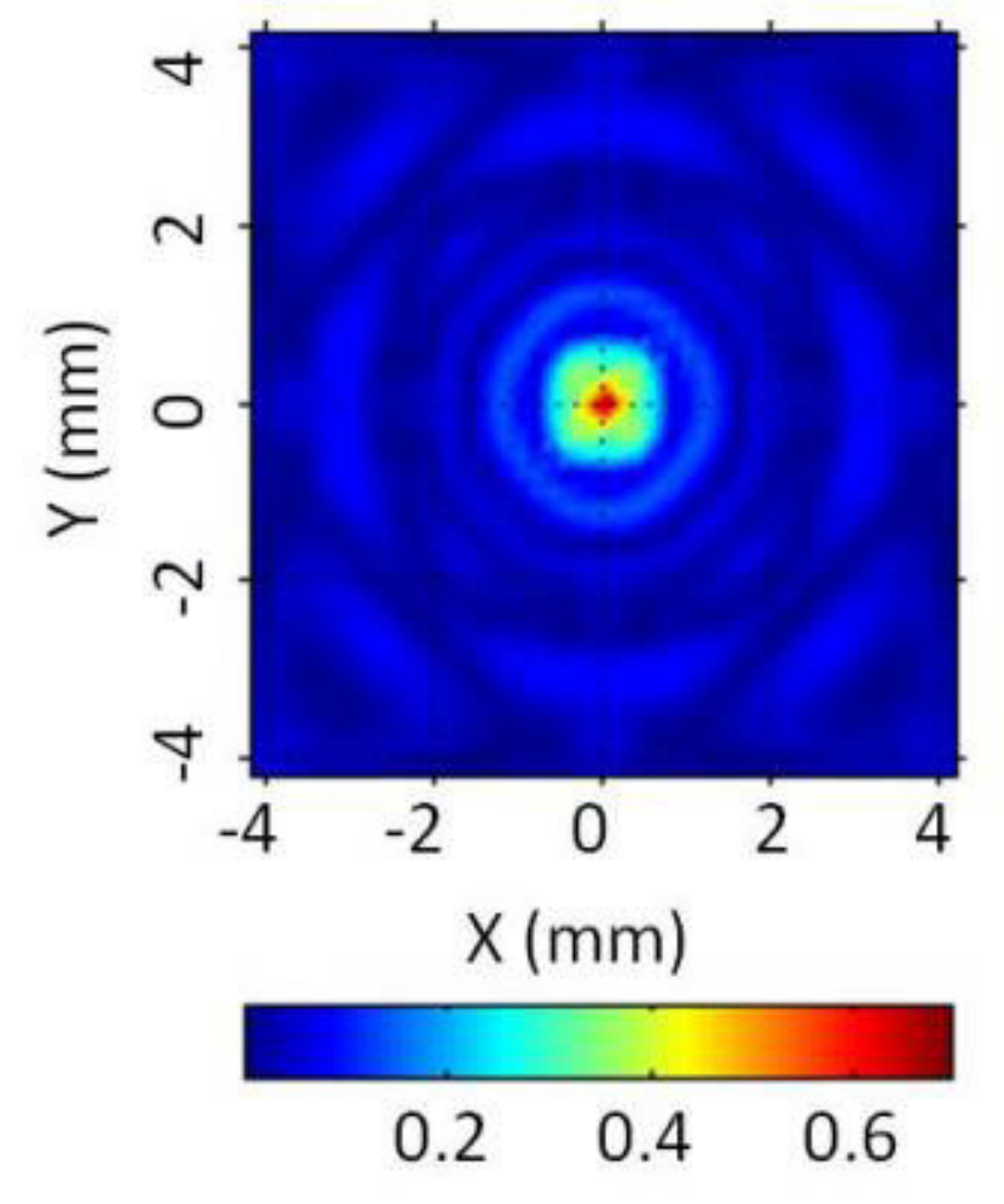} &
\includegraphics[width=0.28\textwidth]{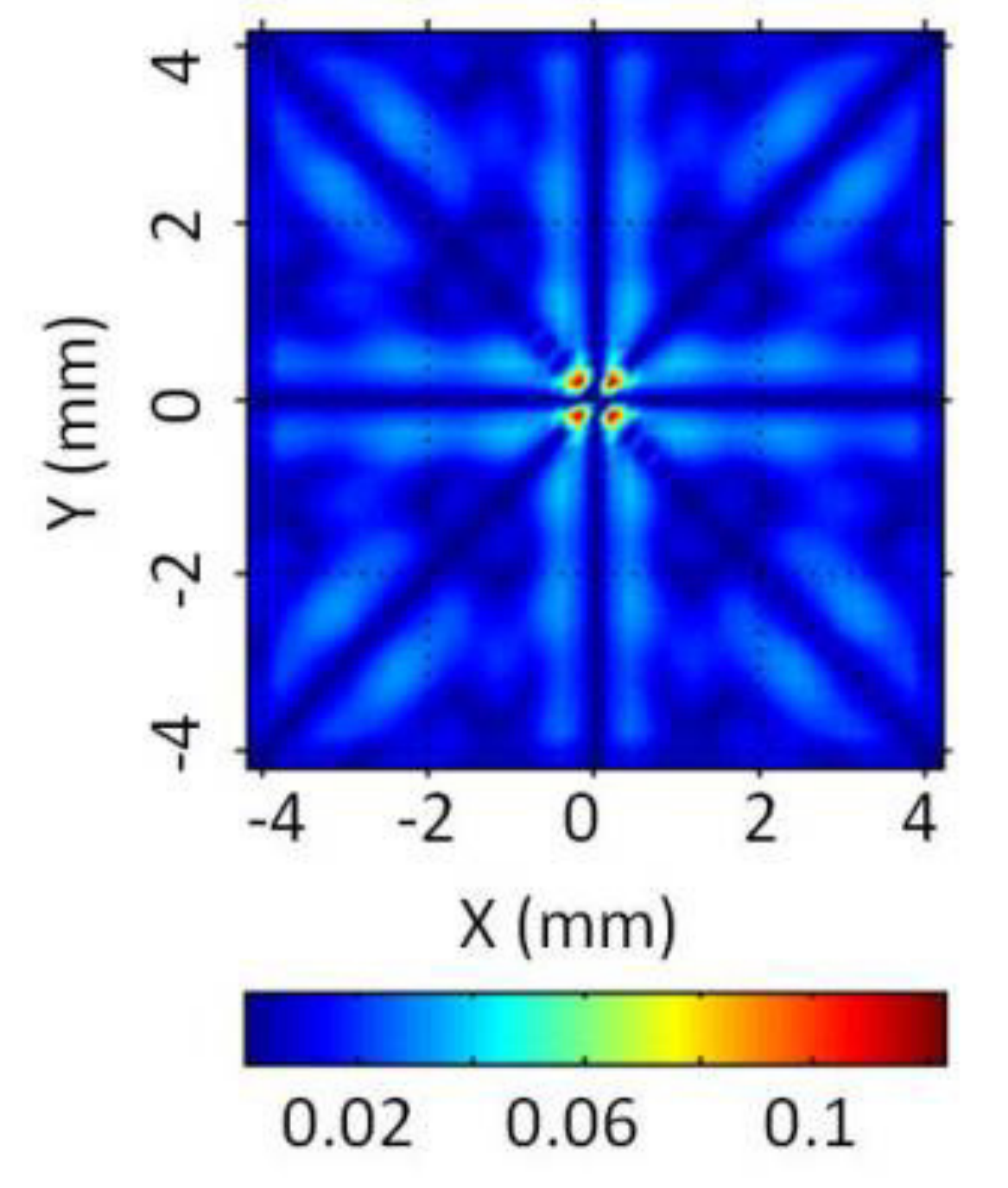}\\
(a) & (b) & (c)\\
\end{tabular}
\end{center}
\vspace{-0.1in}
\caption{Particle displacement components when all four transducers (T1, T2, T3, T4) are active: (a) $u_z$, (b) $u_r$, and (c) $u_\psi$} 
\label{fig:T1234OnOffCases}
\vspace{-0.1in}
\end{figure*}

\item \textit{All four transducers are active}: The case when all four transducers are active is similar to a complete $360^{\circ}$ ring case, except that the four transducers in this case are shifted in space by a small distance (``shift"), which makes it a special case. While the $360^{\circ}$ rings transducer does not generate rotational acoustic effect due to its circular symmetry, the four transducers arranged in an array with a small ``shift" between them produce enough vortexing effect inside the sample. The loci of maximum vortexing field could be shifted away from the center by increasing the shift between the transducers. We can easily notice from figure-\ref{fig:T1234OnOffCases}(a)-(c) that the displacement in the z-direction ($u_z$) and  in the radial directiom ($u_r$) have circular symmetry while the displacement in the azimuth direction ($u_\psi$) has four discreet peak intensity points around the origin shifted by the distance equal to the shift between any two adjacent transducers in the array. Hence we get both converging and vortexing effects from the acoustic field generated from the four sector-transducers arranged in a circular fashion. 

Among all cases, this case provides the highest particle displacement in radial, vertical and rotational directions. Moreover, there are several possible combinations of exciting different number of transducers at a time. Given all this, we selected this case for the DNA shearing experiments.

\end{enumerate}

\begin{figure*}[tb]
\begin{center}
\begin{tabular}{cc}
\includegraphics[width=0.5\textwidth]{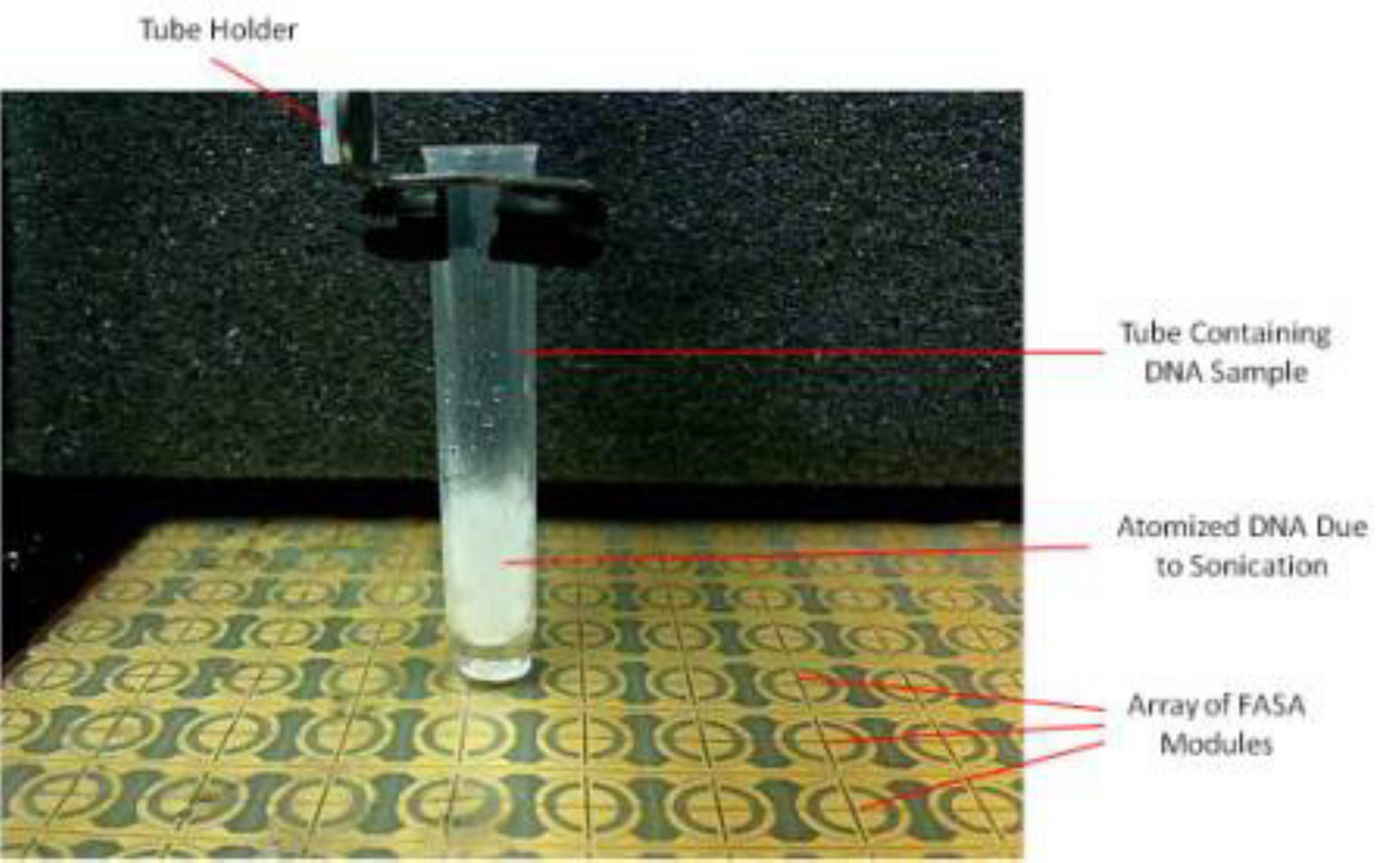} &
\hspace{0.6in}
\includegraphics[width=0.34\textwidth]{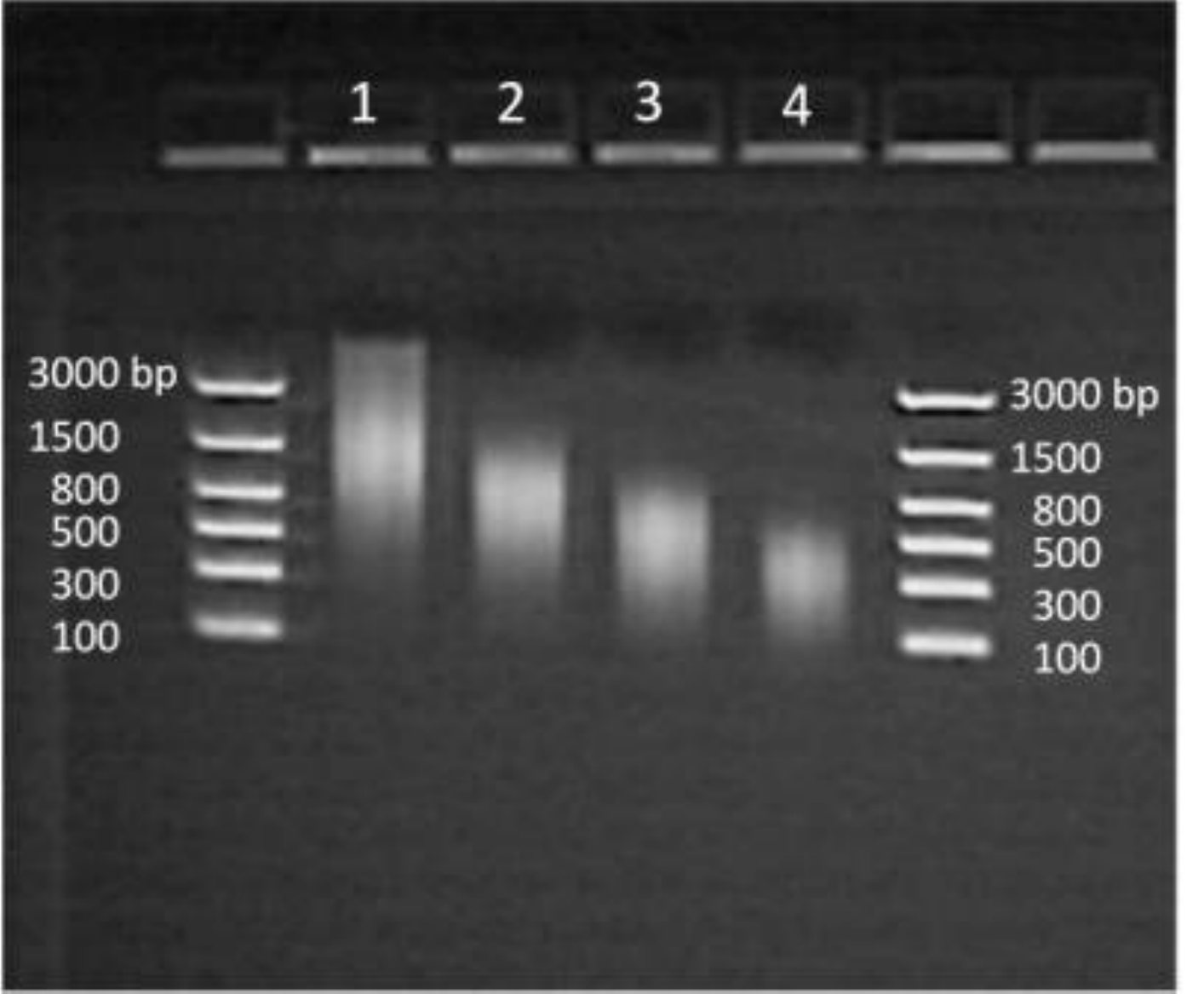}\\
(a) & (b)\\
\end{tabular}
\end{center}
\vspace{-0.1in}
\caption{ (a) Proposed array of phased-array sector-transducer for high-throughput DNA shearing; (b) Gel electrophoresis image for four DNA shearing experiments.} 
\label{fig:Four90DegFASAUrPhiZPerAreaSectoredFASAs}
\vspace{-0.22in}
\end{figure*}

\begin{figure*}[tb]
\begin{center}
\begin{tabular}{cc}
\includegraphics[width=0.42\textwidth]{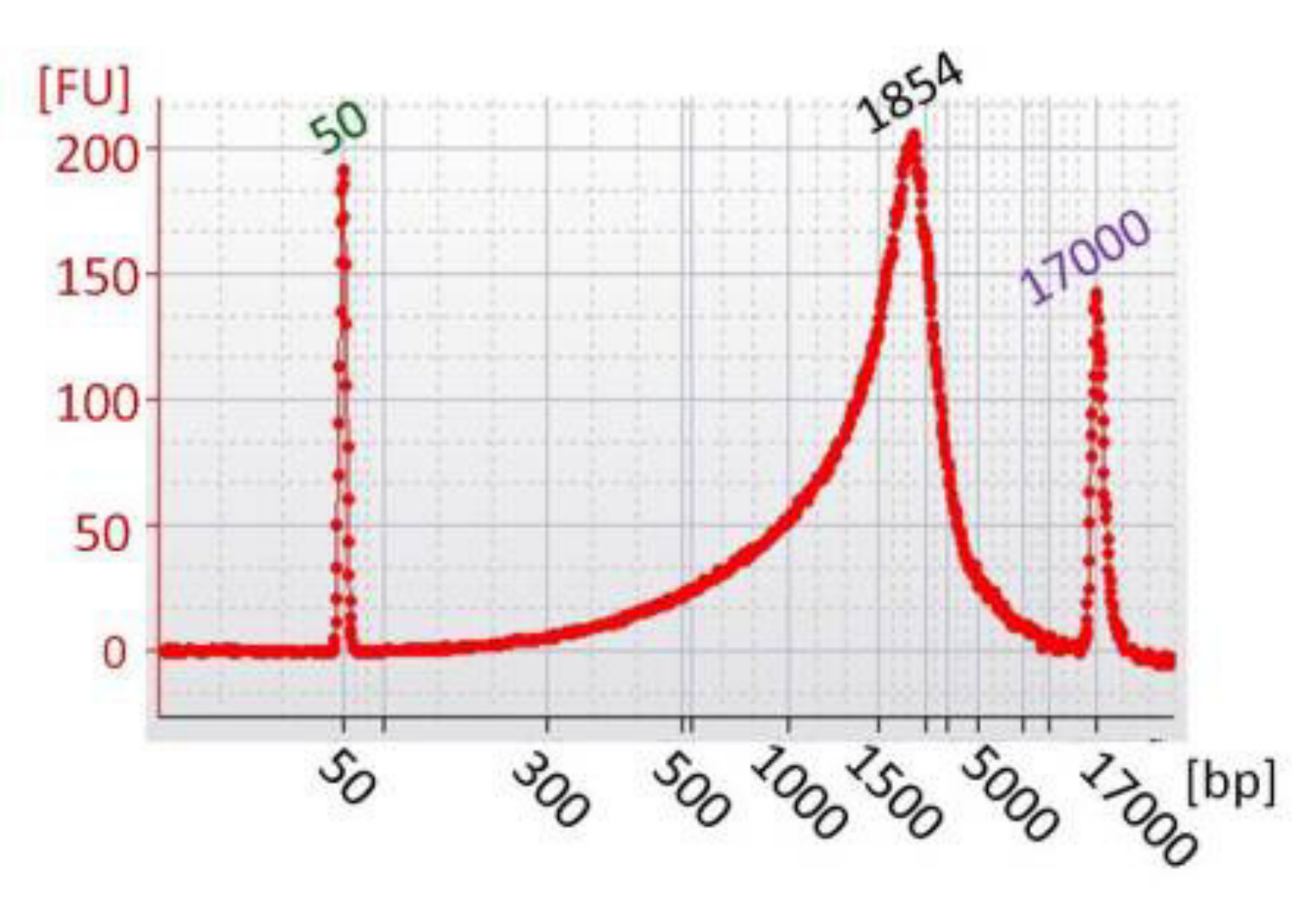} &
\includegraphics[width=0.42\textwidth]{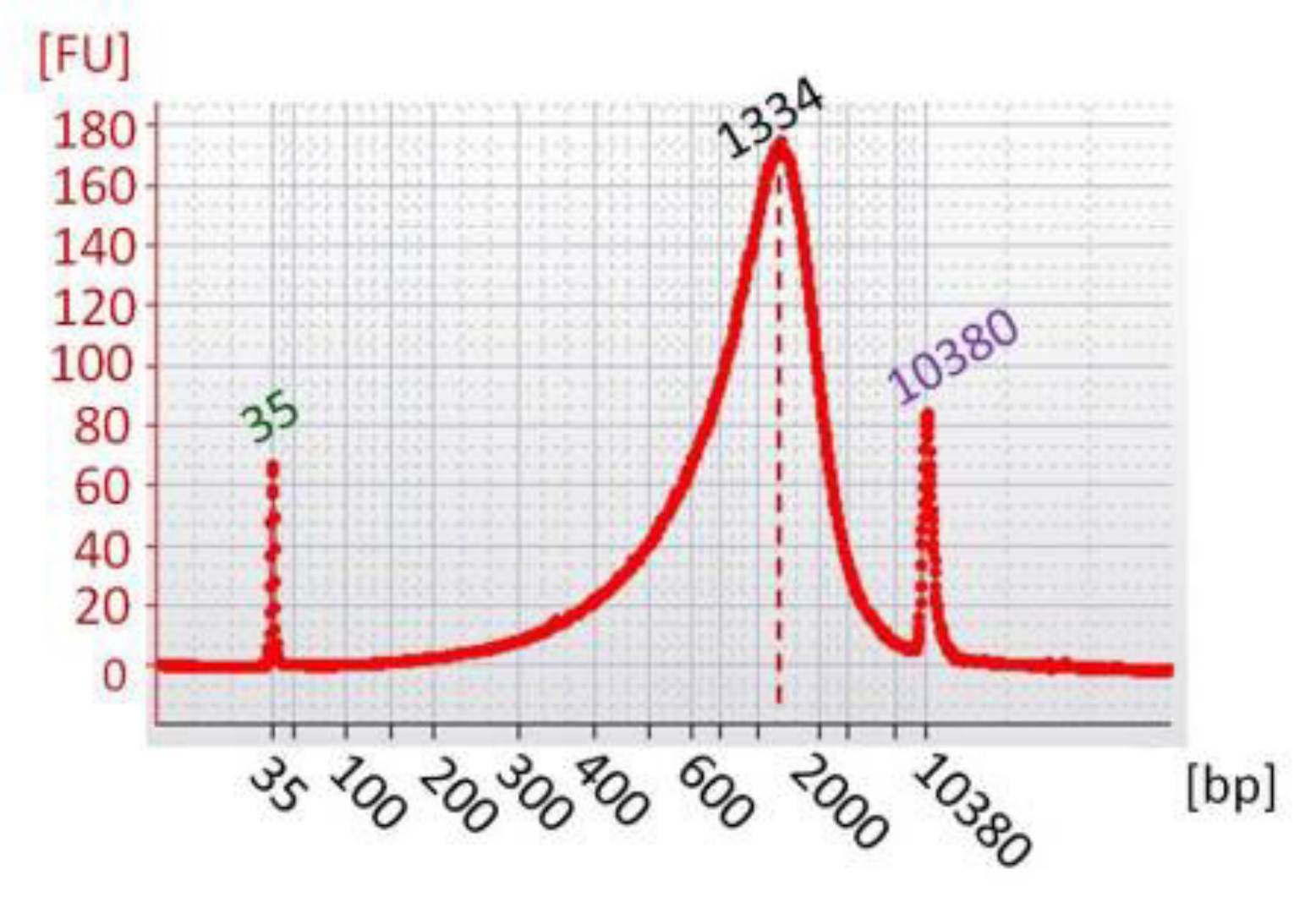}\\
(a) & (b)\\
\includegraphics[width=0.42\textwidth]{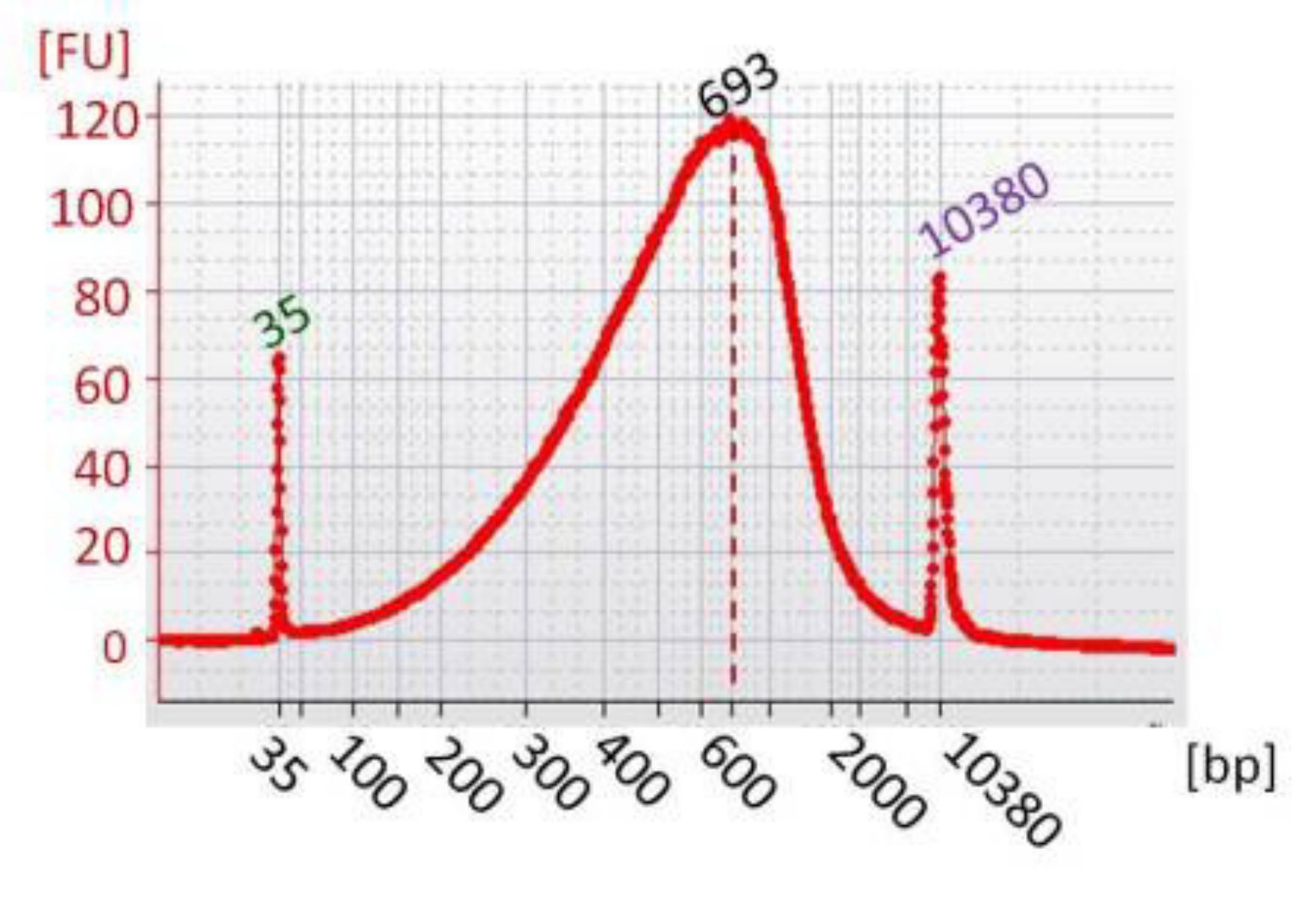} &
\includegraphics[width=0.42\textwidth]{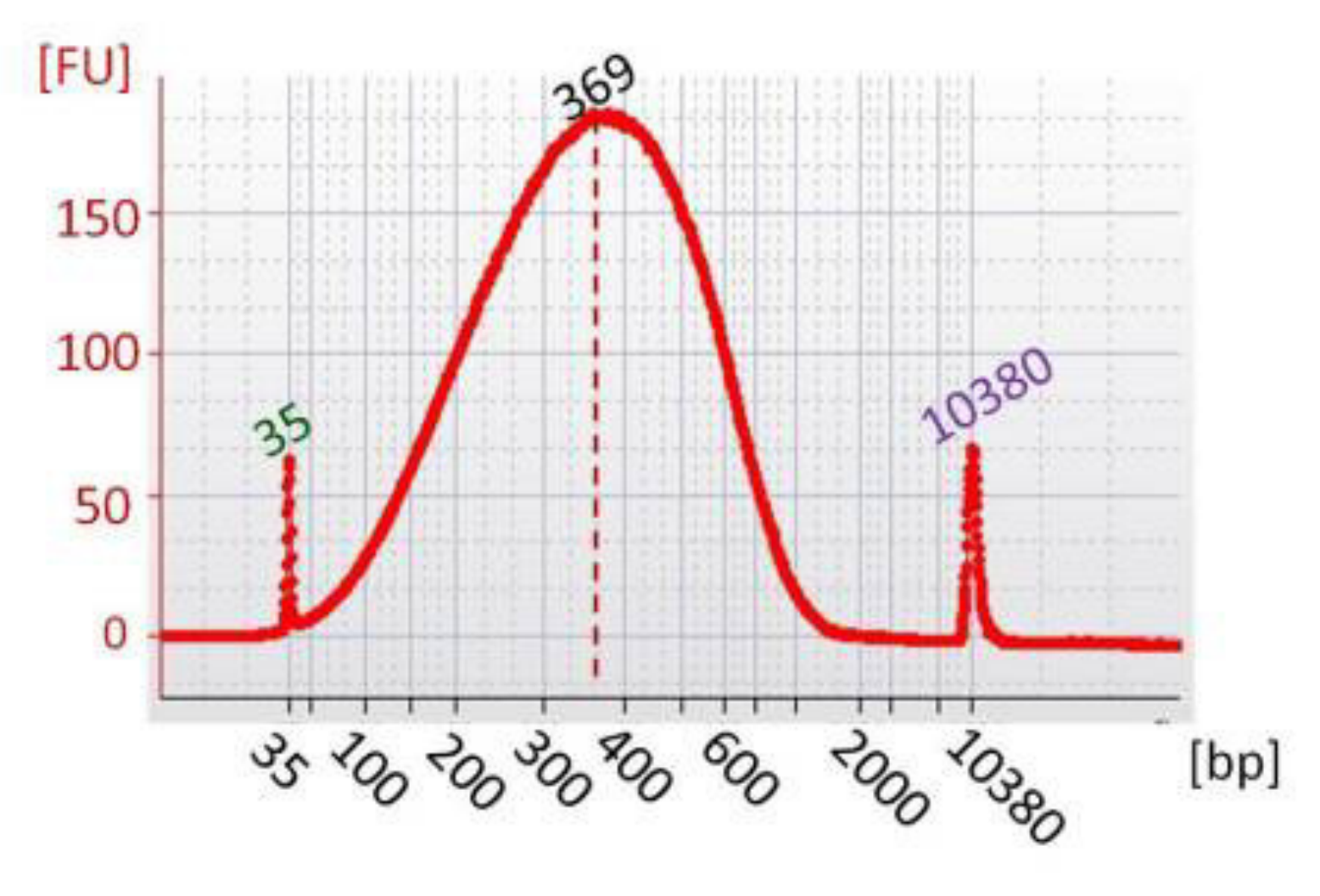}\\
(c) & (d)\\

\end{tabular}
\end{center}
\vspace{-0.1in}
\caption{Agilent Bioanalyzer plots for the four shearing experiments (a) mean fragment size: ~1850; (b) mean fragment size: ~1350; (c) mean fragment size: ~700; (d) mean fagment size: ~370.} 
\label{fig:DNAShearingResults}
\vspace{-0.22in}
\end{figure*}

\section{Experimental Results}
\label{sec:XperimentResults}
The proposed transducer structure is used to shear lambda DNA from E. Coli with 48kbp starting DNA size. The annular electrodes of the transducer were designed to have a focal length of 13 mm and the size of transducer-array is kept as 9mm by 9mm. The RF frequency of operation is decided to be 3.85MHz based on thickness resonant frequency of the transducer. Under these design parameters, each sectored-transducer could only have two rings because the maximum radius of any ring is limited to 4.5mm. The ``shift" between any two adjacent transducers is kept approximately half the wavelength so that four transducers could easily be fit inside 9mm by 9mm.  The proposed array of 96 phased-array sector-transducers for high-throughput DNA shearing is shown in figure-\ref{fig:Four90DegFASAUrPhiZPerAreaSectoredFASAs}-(a).

The small footprint of the transducer array allows us to scale the system for high-throughput DNA shearing required for next generation DNA sequencing instruments. This is one of the limitation of the other sonication based methods presently used by the users. For example, the transducer size used to process one DNA sample using purely focused acoustic waves is in inches, which makes it difficult to scale the system for high-throughput DNA shearing~\cite{Covaris}. Another advantage of the our technology is that the transducer fabrication-process is micro-electromechanical systems (MEMS) based~\cite{MSS}. As the demand of shearing DNA at lower volumes arises, the proposed transducer array could be designed on a lab-on-a-chip device.

Once the transducer design is fixed, the shearing protocol is decided based on following parameters: the starting size of DNA, target mean fragment size, fragment distribution, and volume of DNA. Based on the experiments done on our system, we observed that the peak to peak RF voltage used to excite the transducer, duty cycle of the RF signal, and the sonication time has great effect on the final mean DNA fragment size. We present the results of four experiments in this paper. In the first experiment, we used peak-to-peak RF voltage applied to the transducer array as 140V. The repetition rate of the RF signal is kept at 420Hz for all the experiments. After 15 minutes of sonication at 140V at 1\% duty cycle, the mean size of DNA fragments is found to be about 1854bp. In the second experiment, we increased the voltage to 165V and decreased the duty cycle to 0.75\%; it sheared the lambda DNA to mean fragments of 1350bp size. We further increased the voltage to 190V and 215V and decreased the duty cycle to 0.5\% and 0.25\% respectively; the mean DNA fragment size obtained at these voltage settings are 700bp and 370bp respectively. The sheared DNA were analyzed using both standard gel-electrophoresis and Agilent Bioanalyzer 2100. The gel-image for these four experiments is shown in figure-\ref{fig:Four90DegFASAUrPhiZPerAreaSectoredFASAs}-(b); the image highlights the mark lengths and the distribution of fragmented DNA samples.

To achieve different shearing results, one just needs to find proper experimental settings, also called shearing protocol. A typical shearing protocol comprises following parameters: peak RF voltage, RF frequency (decided by the thickness and material of piezoelectric transducer), repetition rate (duty cycle) of RF signal, phasing between RF signals applied to different transducers in the transducer-array, sonication time, volume of DNA, concentration of DNA, type of DNA, starting size of DNA, target mean size and distribution. The shearing protocols for the four above mentioned experiments are presented in the Table-~\ref{table:ShearingProtocols}. The plots from Agilent Bioanalyzer tool showing peak at the average DNA fragment size are shown in Fig.~\ref{fig:DNAShearingResults}.

\begin{table}[tb]
\caption{Shearing Protocols for Four DNA Shearing Experiments.}
\label{table:ShearingProtocols}
\begin{tabular}{|c|c|c|c|c|}
\hline
Sl. & RF Voltage & Duty Cycle & Sonication & Mean Fragment\\
No. & Vpp (V) & (\%) & Time (mins) & Size (bp)\\
\hline
1 & 140 & 1 & 15 & ~1850\\
\hline
2 & 165 & 0.75 & 15 & ~1350\\
\hline
3 & 190 & 0.5 & 15 & ~700\\
\hline
4 & 215 & 0.25 & 15 & ~370\\
\hline
\end{tabular}
\end{table}

\vspace{-0.1in}

\section{Conclusion}
\label{sec:conclusion}
This paper presents the design of a fast, efficient and controlled DNA shearing system. The system uses an array of four $90^{\circ}$ sectored-transducers, fabricated using MEMS technology, to generate a unique acoustic field pattern, having both converging and vortexing effect inside the DNA sample. The paper provides detailed simulation results for the phased-array transducer used to shear the DNA. In particular, acoustic simulation results for particle displacement are provided for cases when one, two, three, and all four transducers in the array are excited with RF signals.
We also present the shearing protocol used for four DNA shearing experiments performed to demonstrate the shearing capability of the proposed transducer structure. We analyzed the sheared DNA using both gel-electrophoresis and Agilent Bioanalyzer 2100 instrument and confirm that the DNA was sheared to 1850, 1350, 700, and 370 bp when the applied RF voltage was 140V, 165V, 190V, and 215V respectively. The mean fragment size and distribution could be easily changed by changing the shearing protocol. The future work includes the exploration of exciting different number of transducers in the same shearing protocol. 

\bibliographystyle{IEEEtran}
\bibliography{paper_final}
\end{document}